\documentclass[twocolumn,tighten,twocolappendix]{aastex63}
\pdfoutput=1 
\usepackage{amsmath,amstext}
\usepackage[T1]{fontenc}
\usepackage{ae,aecompl}
\usepackage[utf8]{inputenc}
\usepackage{apjfonts} 
\usepackage[figure,figure*]{hypcap}
\usepackage{enumitem}
\usepackage{bm}
\usepackage{graphicx}
\usepackage{lineno}
\usepackage{tikz}

\definecolor{sidm}{HTML}{FF00FF}
\definecolor{wdm_3}{HTML}{FF9966}
\definecolor{wdm_4}{HTML}{FF7733}
\definecolor{wdm_5}{HTML}{FF5500}
\definecolor{wdm_6}{HTML}{CC4400}
\definecolor{wdm_6_5}{HTML}{993300}
\definecolor{wdm_10}{HTML}{662200}

\definecolor{Tkd_072}{HTML}{87CEEB}
\definecolor{Tkd_146}{HTML}{1E90FF}
\definecolor{Tkd_232}{HTML}{191970}

\input{hyperlink-year-only-natbib-patch}

\newcommand*{\http}[1]{\href{http://#1}{#1}}
\newcommand*{\https}[1]{\href{https://#1}{#1}}

\shorttitle{COZMIC. III.}
\shortauthors{Nadler et al.}

\begin{document}

\title{COZMIC. III. Cosmological Zoom-in Simulations of Self-interacting Dark Matter with Suppressed Initial Conditions}

\author[0000-0002-1182-3825]{Ethan O.~Nadler}
\affiliation{Department of Astronomy \& Astrophysics, University of California, San Diego, La Jolla, CA 92093, USA}
\affiliation{Carnegie Observatories, 813 Santa Barbara Street, Pasadena, CA 91101, USA}
\affiliation{Department of Physics $\&$ Astronomy, University of Southern California, Los Angeles, CA 90007, USA}

\author[0000-0001-9543-5012]{Rui An}
\affiliation{Department of Physics $\&$ Astronomy, University of Southern California, Los Angeles, CA 90007, USA}

\author[0000-0002-5421-3138]{Daneng Yang}
\affiliation{Purple Mountain Observatory, Chinese Academy of Sciences, Nanjing 210023, China}
\affiliation{Department of Physics and Astronomy, University of California, Riverside, CA 92521, USA}

\author[0000-0002-8421-8597]{Hai-Bo Yu}
\affiliation{Department of Physics and Astronomy, University of California, Riverside, CA 92521, USA}

\author[0000-0001-5501-6008]{Andrew Benson}
\affiliation{Carnegie Observatories, 813 Santa Barbara Street, Pasadena, CA 91101, USA}

\author[0000-0002-3589-8637]{Vera Gluscevic}
\affiliation{Department of Physics $\&$ Astronomy, University of Southern California, Los Angeles, CA 90007, USA}

\correspondingauthor{Ethan~O.~Nadler}
\email{enadler@ucsd.edu}

\label{firstpage}

\begin{abstract}
We present eight cosmological dark matter (DM)--only zoom-in simulations of a Milky Way--like system that include suppression of the linear matter power spectrum $P(k)$, and/or velocity-dependent DM self-interactions, as the third installment of the COZMIC suite. We consider a model featuring a massive dark photon that mediates DM self-interactions and decays into massless dark fermions. The dark photon and dark fermions suppress linear matter perturbations, resulting in dark acoustic oscillations in $P(k)$, which ultimately affect dwarf galaxy scales. The model also features a velocity-dependent elastic self-interaction between DM particles (SIDM), with a cross section that can alleviate small-scale structure anomalies. For the first time, our simulations test the impact of $P(k)$ suppression on gravothermal evolution in an SIDM scenario that leads to core collapse in (sub)halos with present-day virial masses below $\approx 10^9~M_{\mathrm{\odot}}$. In simulations with $P(k)$ suppression and self-interactions, the lack of low-mass (sub)halos and the delayed growth of structure reduce the fraction of core-collapsed systems relative to SIDM simulations without $P(k)$ suppression. In particular, $P(k)$ suppression that saturates current warm DM constraints almost entirely erases core collapse in isolated halos. Models with less extreme $P(k)$ suppression produce core collapse in $\approx 20\%$ of subhalos and $\approx 5\%$ of isolated halos above $10^8~M_{\mathrm{\odot}}$, and also increase the abundance of extremely low-concentration isolated low-mass halos relative to SIDM. These results reveal a complex interplay between early and late-Universe DM physics, revealing new discovery scenarios in the context of upcoming small-scale structure measurements.
\end{abstract}

\keywords{\href{http://astrothesaurus.org/uat/353}{Dark matter (353)}; 
\href{http://astrothesaurus.org/uat/1083}{$N$-body simulations (1083)};
\href{http://astrothesaurus.org/uat/1880}{Galaxy dark matter halos (1880)}}

\section{Introduction}\label{sec:intro}

Small-scale cosmic structure is a key probe of dark matter (DM) physics, including its interactions (e.g., see \citealt{Buckley171200615} and \citealt{Gluscevic190305140} for reviews). Different kinds of DM physics beyond gravity---i.e., beyond the cold, collisionless DM (CDM) paradigm---are often parameterized and modeled separately. For example, free-streaming of warm DM particles (WDM; \citealt{Bond1983,Bode0010389}) imprints a small-scale cutoff in the linear matter power spectrum $P(k)$, which reduces low-mass halo and subhalo abundances relative to CDM. Meanwhile, interactions within the dark sector, captured by self-interacting DM (SIDM) models (\citealt{Spergel9909386}), affect (sub)halos' internal structure, including their density profiles and shapes. These signatures are often intertwined; for example, SIDM can suppress subhalo abundances through subhalo--host halo interactions and enhanced stripping of cored halos (e.g., \citealt{Vogelsberger12015892,Nadler200108754}).

$P(k)$ suppression associated with WDM--like models is not typically included when modeling SIDM. However, models that feature DM interactions often affect $P(k)$. This is well known for DM that interacts with the standard model or dark radiation (e.g., \citealt{Boehm0012504,Boehm0410591,Feng09053039,vandenAarssen12055908,Buckley14052075}). For example, \cite{Boehm0112522} show that DM--standard model interactions can impact $P(k)$ in a way that mimics WDM (also see \citealt{Nadler190410000}). On the other hand, DM self-interactions are inefficient until the high densities at the centers of virialized halos are achieved. For this reason, SIDM does not significantly alter large-scale structure observables relative to CDM (e.g., \citealt{Rocha12083025}).

Although the direct impact of self-interactions on $P(k)$ is minor, favored SIDM models feature a velocity-dependent cross section with a low-mass dark mediator (e.g., see \citealt{Tulin170502358} and \citealt{Adhikari220710638} for reviews); this mediator, and other dark sector particles that accompany the self-interactions, affect the growth of linear matter perturbations and suppress $P(k)$. These scenarios therefore lead to WDM--like initial conditions (ICs), but also feature SIDM phenomenology at late times; thus we refer to them as WSIDM.

As a benchmark example of WSIDM, we consider a model with a $0.1~\mathrm{GeV}$ DM particle and an $\sim 10~\mathrm{keV}$ dark photon that mediates DM self-interactions \citep{Huo170909717}. The mediator couples to a dark fermion, which increases the effective number of relativistic species at early times. We will show that (sub)halo abundances and density profiles are sensitive to both $P(k)$ suppression and self-interactions. Crucially, our WSIDM scenario features a velocity-dependent SIDM cross section that can alleviate small-scale structure anomalies by producing core collapse in low-mass (sub)halos \citep{Balberg0110561,Koda11013097,Essig180901144}. Furthermore, it produces $P(k)$ cutoffs tested by current small-scale structure data, allowing us to assess detectability of WSIDM with upcoming data. 

We note that the same combination of effects arises in the effective theory of structure formation (ETHOS) framework (\citealt{Cyr-Racine151205344,Vogelsberger151205349}). However, as we will demonstrate, $P(k)$ suppression is so strong in \cite{Vogelsberger151205349} that core collapse is not achieved. Thus, to date, no WSIDM simulations have been performed which include $P(k)$ suppression and capture the full range of (sub)halos' gravothermal evolution. It is critical to understand the interplay between these effects.

\begin{figure*}[t!]
\centering
\includegraphics[trim={0 0cm 0 0cm},width=\textwidth]{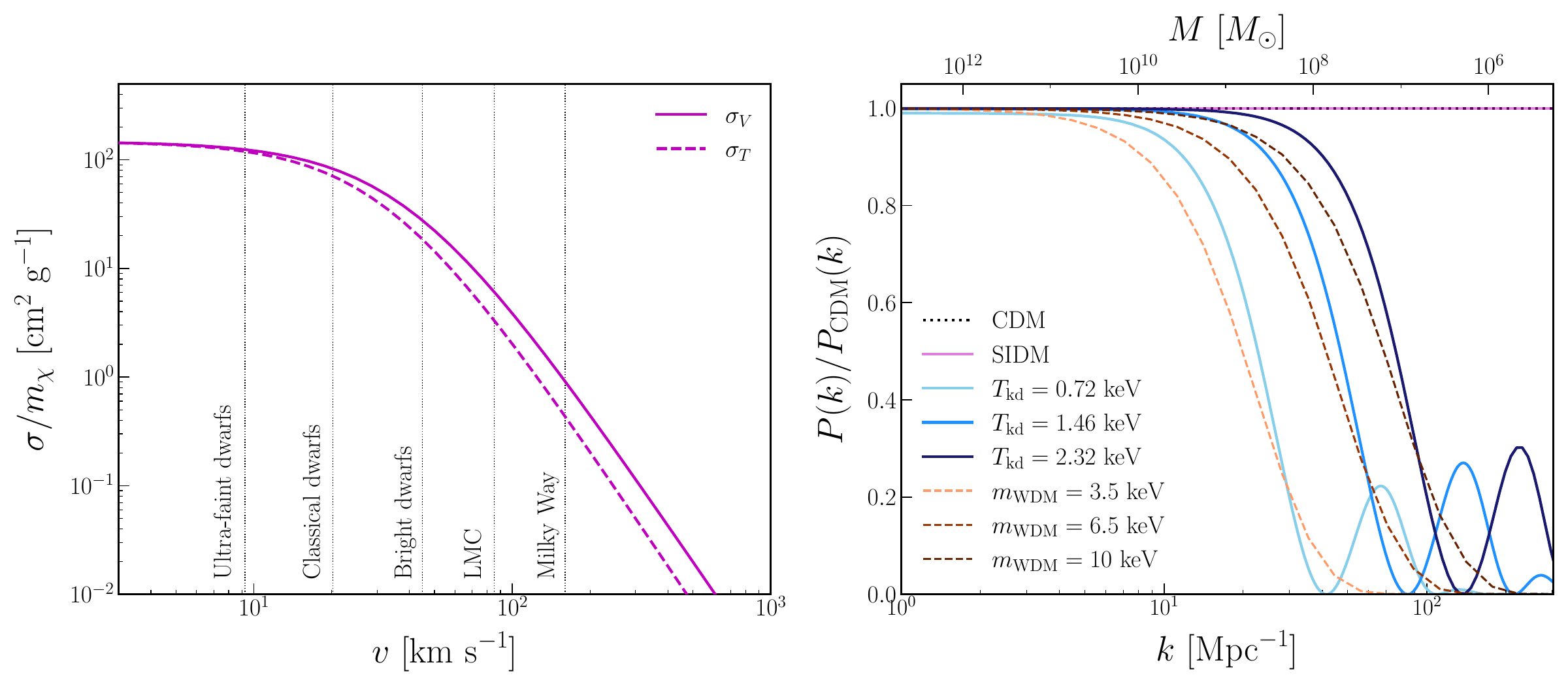}
    \caption{Left panel: viscosity (solid) and momentum-transfer (dashed) self-interaction cross sections, which define our MilkyWaySIDM model \citep{Yang221113768}. Note that our SIDM implementation captures the velocity and angular dependence of the differential cross section (see Section~\ref{sec:zoom-ins}). Vertical dotted lines show maximum circular velocities of our MW--like host ($M_{\mathrm{vir}}=10^{12}~M_{\mathrm{\odot}}$), LMC analog ($M_{\mathrm{vir}}=10^{11}~M_{\mathrm{\odot}}$), and halos expected to host bright ($M_{\mathrm{vir}}=10^{10}~M_{\mathrm{\odot}}$), classical ($M_{\mathrm{vir}}=10^9~M_{\mathrm{\odot}}$), and ultrafaint ($M_{\mathrm{vir}}=10^8~M_{\mathrm{\odot}}$) dwarf galaxies. Right panel: ratio of the linear matter power spectrum for each model we simulate (colored lines) relative to CDM (horizontal dotted black line). Our SIDM simulation (magenta) uses CDM ICs. We simulate $T_{\mathrm{kd}}=0.72$, $1.46$, and $2.32~\mathrm{keV}$ (lightest to darkest blue) with and without self-interactions; we simulate WDM models with $m_{\mathrm{WDM}}=3.5$, $6.5$, and $10~\mathrm{keV}$ without self-interactions (lightest to darkest red). Top ticks show halo masses calculated in linear theory (\citealt{Nadler190410000}; \citetalias{Nadler241003635}). The largest scale shown roughly corresponds to the size of our zoom-in regions, and the smallest to the Nyquist frequency of the high-resolution simulations used for our main analyses. }
    \label{fig:transfer}
\end{figure*}

Here, we perform the first WSIDM simulations that capture core collapse. The SIDM cross section we consider (hereafter ``MilkyWaySIDM'') was originally presented in \cite{Yang221113768} and is shown in the left panel of Figure~\ref{fig:transfer}. This model is similar to those considered in \cite{Correa200702958} and \cite{Turner201002924}, and is motivated as follows. At large velocities, $v\approx 1000~\mathrm{km\ s}^{-1}$, observations of galaxy clusters and large elliptical galaxies place stringent upper limits on the SIDM cross section of $\approx 0.1$ to $1~\mathrm{cm}^2~\mathrm{g}^{-1}$ \citep{Kaplinghat150803339,Sagunski200612515,Andrade201206611,Kong240215840}. Constraints are weaker at lower velocities, corresponding to galaxy group and MW scales. At dwarf galaxy scales ($v\approx 10~\mathrm{km\ s}^{-1}$) data rule out a velocity-independent cross section above $\approx 5~\mathrm{cm}^2~\mathrm{g}^{-1}$ (e.g., \citealt{Correa200702958,Silverman220310104}). In particular, a velocity-independent cross section on dwarf scales predicts uniformly cored subhalos, which are challenging to reconcile with MW satellites' observed inner densities and bright dwarfs' diverse rotation curves (e.g., \citealt{Read180506934,Valli171103502,Zavala190409998,Kim210609050,Silverman220310104,Yang221113768}). This challenge is compounded by the orbital properties of observed satellites (e.g., \citealt{Kaplinghat190404939,Sameie190407872,Slone210803243}).

On the other hand, the MilkyWaySIDM cross section predicts diverse (sub)halo density profiles due to core formation and collapse. Strong, velocity-dependent SIDM models can explain observations of very low-concentration ultradiffuse galaxies (e.g., \citealt{Kong220405981,Zhang240104985}), extremely concentrated strong-lensing substructure detected by gravitational imaging \citep{Minor201110627,Yang210202375,Nadler230601830,Zeng231009910}, and the properties of the GD-1 stellar stream perturber \citep{Zhang240919493}. These scenarios can also be constrained with strong-lensing flux ratio statistics \citep{Gilman210505259,Gilman220713111,Tran240502388}.

Previous studies that simulated the MilkyWaySIDM cross section or similar models all assumed CDM initial conditions (ICs), with the exception of \cite{Vogelsberger151205349}. The characteristic suppression scale in $P(k)$ is determined by the kinetic decoupling temperature of the dark fermion--DM interaction, $T_{\mathrm{kd}}$. In our scenario, $T_{\mathrm{kd}}$ is controlled by the DM self-coupling, mediator mass, and dark fermion temperature \citep{Huo170909717}. We will show that natural values for these parameters yield a $P(k)$ cutoff on dwarf galaxy scales, thereby affecting the formation and evolution of the same (sub)halos that are impacted by self-interactions. In particular, we consider three $P(k)$ cutoffs that are constrained by current observations of the Ly$\alpha$ forest (\citealt{Viel1308804,Irsic179602}, \citeyear{Irsic230904533}), MW satellites \citep{Nadler200800022}, and strong gravitational lensing \citep{Nadler210107810}. The corresponding transfer functions set the ICs for our simulations and are shown in the right panel of Figure~\ref{fig:transfer}.

With these ingredients, we run eight new high-resolution cosmological DM--only zoom-in simulations of an MW analog from the Milky Way-est suite \citep{Buch240408043}, in WSIDM models.\footnote{We use ``SIDM'' for models with only self-interactions and no $P(k)$ suppression, i.e., SIDM that is cold on all scales relevant for our simulations.} Our simulations are part of the \textbf{CO}smological \textbf{Z}oo\textbf{M}-in simulations with \textbf{I}nitial \textbf{C}onditions beyond CDM (COZMIC) suite, which includes over $100$ beyond-CDM simulations with ICs for warm, fuzzy, and baryon-scattering DM \defcitealias{Nadler241003635}{Paper~I}(Nadler et al.\ \citeyear{Nadler241003635}, hereafter \citetalias{Nadler241003635}), and for models with a fractional non-CDM component (Paper II; \citealt{An241103431}). We show that the severity of $P(k)$ suppression influences whether SIDM core collapse occurs or not. Thus, WSIDM can change predictions for SIDM (sub)halo profiles and simultaneously affect (sub)halo abundances. Combining upcoming observations of strong gravitational lenses, to probe (sub)halo density profiles \citep{Minor201110629,Despali240712910}, and the MW satellite population, to probe the subhalo mass function (SHMF; \citealt{Nadler240110318}), will therefore test WSIDM.

This paper is organized as follows. In Section~\ref{sec:model}, we describe our WSIDM model and how it maps to both $P(k)$ and the SIDM cross section. In Section~\ref{sec:simulations}, we describe our method for generating WSIDM ICs and performing zoom-in simulations that include both suppressed ICs and self-interactions. In Section~\ref{sec:results}, we present our results, focusing on (S)HMFs and structural properties. In Section~\ref{sec:f_cc}, we study how $P(k)$ suppression and gravothermal evolution interplay using matched (sub)halos from our simulations; we make predictions for dwarf galaxy density profiles in Section~\ref{sec:dwarf}. We discuss our results in Section~\ref{sec:discussion} and conclude in Section~\ref{sec:conclusions}.

We adopt the following cosmological parameters: $h = 0.7$, $\Omega_{\rm m} = 0.286$, $\Omega_b = 0.049$, $\Omega_{\Lambda} = 0.714$, $\sigma_8 = 0.82$, and $n_s=0.96$ (\citetalias{Nadler241003635}; \citealt{Hinshaw_2013}). Halo masses are defined via the \cite{Bryan_1998} virial overdensity, which corresponds to $\Delta_{\mathrm{vir}}\approx 99\times \rho_{\mathrm{crit}}$ in our cosmology, where $\rho_{\mathrm{crit}}$ is the critical density of the Universe at $z=0$. We refer to halos within the virial radius of our MW host as ``subhalos,'' and halos outside the virial radius of any larger halo as ``isolated halos.'' We work in natural units with $c=\hbar=1$.


\section{Warm Self-interacting Dark Matter Model}
\label{sec:model}

We consider the interaction Lagrangian \citep{Huo170909717}
\begin{equation}
    \mathcal{L}_{\mathrm{int}} = -ig_{\chi}\bar{\chi}\gamma^{\mu}\chi\phi_{\mu} + m_{\chi}\bar{\chi}\chi + \frac{1}{2}m_{\phi}^2\phi^{\mu}\phi_{\mu} - ig_{f}\bar{f}\gamma^{\mu}f\phi_{\mu},\label{eq:lagrangian}
\end{equation}
which contains an SIDM particle $\chi$ with mass $m_{\chi}$, a massless dark fermion $f$, and an SIDM mediator $\phi$ with mass $m_{\phi}$. The coupling constants $g_{\chi}$ ($g_f$) determine the strength of DM self-interactions (dark fermion--mediator interactions); we follow \cite{Huo170909717} and set $g_{\chi}=g_f$. Finally, $\xi$ parameterizes the ratio of dark-to-visible sector temperature \citep{Feng08082318}. In this scenario, the DM relic abundance is determined by the annihilation of $\chi$ to $\phi$ in the early Universe. Matching $\Omega_{\mathrm{m}}$ to the DM relic density thus imposes a relation between $g_{\chi}$ and $m_{\chi}$. Here, we fix $m_{\chi}=0.1~\mathrm{GeV}$, which determines both $g_{\chi}$ and $m_{\phi}$ based on the desired SIDM cross section, as described below. We therefore remain agnostic about the details of the underlying DM production mechanism.

Meanwhile, the dark fermion contributes to the effective number of relativistic degrees of freedom in the early Universe, $N_{\mathrm{eff}}$, with $N_{\mathrm{eff}}=3.046+\Delta N_{\mathrm{eff}}=3.046+(11/4)^{4/3}\xi^4$. With $m_{\chi}$ fixed and $g_{\chi}$ and $m_{\phi}$ chosen to yield our desired SIDM cross section, $\xi$ is the only remaining parameter that determines the cutoff scale in $P(k)$. Below, we derive three values of $\xi$ that produce desired $P(k)$ cutoff scales; all of these are consistent with bounds from Planck on $\Delta N_{\mathrm{eff}}$, which imply $\xi\lesssim 0.5$ \citep{Archidiacono170606870}. Thus, our model yields a DM relic abundance and expansion history consistent with observations while producing a specific SIDM cross section and desired $P(k)$ cutoffs.

Specifically, to derive $g_{\chi}$ and $m_{\phi}$, we write the differential SIDM cross section as \citep{Yang220503392}
\begin{equation}
    \frac{\mathrm{d}\sigma}{\mathrm{d}\cos\theta} = \frac{\sigma_0w^4}{2\left[w^2+v^2\sin^2(\theta/2)\right]^2},\label{eq:xsec}
\end{equation}
where $\sigma_0\equiv g_{\chi}^4/(4\pi m_{\chi}^2 w^4)$ is the cross-section amplitude, $w\equiv m_{\phi}/m_{\chi}$, $v$ is the relative velocity of SIDM particles, and $\theta$ is their scattering angle. We use the MilkyWaySIDM model from \cite{Yang221113768}, which is similar to the models proposed in \cite{Correa200702958} and \cite{Turner201002924}. In particular, we set $\sigma_0=147.1~\mathrm{cm^2~g^{-1}}$ and $w=24.33~\mathrm{km\ s}^{-1}$, which yields $g_{\chi}=7.8\times 10^{-4}$ and $m_{\phi}=8.11~\mathrm{keV}$ given $m_{\chi}=0.1~\mathrm{GeV}$. Note that $g_{\chi}^2/w \ll 1$, so the Born approximation for $\mathrm{d}\sigma/\mathrm{d}\cos\theta$ is valid \citep{Feng09110422,Tulin13023898}. The left panel of Figure~\ref{fig:transfer} shows the momentum-transfer and viscosity cross sections, calculated following \cite{Yang220503392}, \cite{Yang221113768}, and \cite{Nadler230601830}. Note that the viscosity cross section effectively captures the full, angularly dependent SIDM scattering \citep{Yang220503392}.

The impact of this WSIDM model on $P(k)$ is parameterized by the temperature of the visible sector $T_{\mathrm{kd}}$ when DM kinetically decouples with the dark fermion, given by \citep{Huo170909717}
\begin{align}
    T_{\mathrm{kd}} &= \frac{1.4~\mathrm{keV}}{\sqrt{g_{\chi}g_f}}\left(\frac{m_{\chi}}{100~\mathrm{GeV}}\right)^{\frac{1}{4}}\left(\frac{m_{\phi}}{10~\mathrm{MeV}}\right)\left(\frac{g_{*}}{3.38}\right)^{\frac{1}{8}}\left(\frac{\xi}{0.5}\right)^{-\frac{3}{2}}&\nonumber \\ &= 0.26~\mathrm{keV}\times \left(\frac{\xi}{0.5}\right)^{-\frac{3}{2}},&\label{eq:tkd}
\end{align}
where $g_{*}$ is the number of relativistic degrees of freedom at the redshift of decoupling, which we always set to $3.38$, as kinetic decoupling occurs after electron--positron annihilation in all models we consider. We choose values of $\xi$ such that the resulting half-mode wavenumber $k_{\mathrm{hm}}$, defined by $P(k_{\mathrm{hm}})/P_{\mathrm{CDM}}(k_{\mathrm{hm}})\equiv 0.25$ and returned by our $P(k)$ calculations in Section~\ref{sec:ics}, matches three benchmark models of thermal-relic WDM: $m_{\mathrm{WDM}}=3.5$, $6.5$, and $10~\mathrm{keV}$. This yields $T_{\mathrm{kd}}=0.72$, $1.46$, and $2.32~\mathrm{keV}$, respectively. These correspond to $\xi = 0.25$, $0.15$, and $0.11$, all of which are consistent with the Planck constraint mentioned above.

The right panel of Figure~\ref{fig:transfer} shows linear matter power spectra normalized to CDM for these scenarios, calculated using \textsc{CLASS}. The cutoff scales are respectively chosen to saturate current WDM limits from the Ly$\alpha$ forest (\citealt{Viel1308804,Irsic179602}, \citeyear{Irsic230904533}), the MW satellite population \citep{Nadler200800022}, and a combination of MW satellites and strong gravitational lensing flux ratio statistics \citep{Nadler210107810}. The $10~\mathrm{keV}$ case is also comparable to the WDM sensitivity of upcoming dwarf galaxy surveys \citep{Nadler240110318}. Our most suppressed scenario, $T_{\mathrm{kd}}=0.72~\mathrm{keV}$, is slightly colder than the ETHOS-2 ($T_{\mathrm{kd}}=0.33~\mathrm{keV}$) and ETHOS-3 ($T_{\mathrm{kd}}=0.51~\mathrm{keV}$) models from \cite{Vogelsberger151205349} and \cite{Sameie181011040}. The other models have cutoffs on smaller scales and are less suppressed than the $T_{\mathrm{kd}}=1~\mathrm{keV}$ model studied in \cite{Huo170909717}.


\section{Simulations}
\label{sec:simulations}

We now describe our procedure for generating ICs (Section~\ref{sec:ics}) and running our zoom-in simulations (Section~\ref{sec:zoom-ins}). 

\subsection{Initial Conditions}
\label{sec:ics}

To generate transfer functions for our WSIDM zoom-in simulations, we run a modified version of \textsc{CAMB} \citep{Lewis0205436} developed by \cite{Cyr-Racine151205344}. We also generate transfer functions for the benchmark WDM models discussed above using \textsc{CLASS} (\citealt{class}; see Paper I for details). The right panel of Figure~\ref{fig:transfer} compares all of the transfer functions we simulate. In addition to the initial cutoff in $P(k)$, our $T_{\mathrm{kd}}$ models feature dark acoustic oscillations (DAOs) on very small scales. Based on the results from Paper I, we do not expect these low-amplitude DAOs to significantly alter (sub)halo abundances compared to the half-mode-matched WDM models; we confirm and quantify this statement in Section~\ref{sec:shmf} and Appendix~\ref{sec:wdm_comparison}.

Following Paper I, we input these transfer functions to \textsc{MUSIC} \citep{Hahn11036031} to generate ICs for one of the MW--like hosts from \citetalias{Nadler241003635}, Halo004; CDM ICs were originally drawn from the Milky Way-est suite \citep{Buch240408043}. This host has a $z=0$ virial mass of $1.03\times 10^{12}~M_{\mathrm{\odot}}$, contains an LMC analog subhalo that accreted $\approx 1~\mathrm{Gyr}$ before $z=0$ with a peak virial mass of $1.74\times 10^{11}~M_{\mathrm{\odot}}$, and merges with an analog of Gaia--Sausage--Enceladus at $z=1.4$ (\citealt{Buch240408043}; \citetalias{Nadler241003635}). For our main analyses, we focus on ``high-resolution'' simulations initialized at $z=99$ using five refinement regions, with a DM particle mass of $m_{\mathrm{part}}=5.0\times 10^4~M_{\mathrm{\odot}}$ and Plummer-equivalent gravitational softening of $\epsilon=80~\mathrm{pc}~h^{-1}$ in the highest-resolution region, which corresponds to the Lagrangian volume of particles within $10$ times
the virial radius $R_{\mathrm{vir}}$ of the MW host in the parent box at $z=0$. In Appendix~\ref{sec:convergence}, we simulate the same models with one fewer refinement region to assess convergence.

\subsection{Zoom-in Simulations}
\label{sec:zoom-ins}

We run COZMIC III simulations using a version of \textsc{Gadget-2} \citep{Springel0505010} modified to include SIDM, with other code settings identical to \citetalias{Nadler241003635}. We use the SIDM implementation from \cite{Yang221113768}, with elastic self-interactions modeled as described in \cite{Yang220503392}. This implementation captures the full velocity and angular dependence of the differential SIDM scattering cross section (Equation~\ref{eq:xsec}). We generate halo catalogs and merger trees using {\sc Rockstar} and {\sc consistent-trees} \citep{Behroozi11104372,Behroozi11104370}, and we analyze all simulations at $z=0$. We analyze all subhalos within the virial radius of the MW host and all isolated halos within $3~\mathrm{Mpc}$ (or $\approx 10R_{\mathrm{vir}}$) of the MW center, following \cite{Yang221113768}; the fraction of low-resolution (LR) particles within this volume is negligible in all cases.

We simulate CDM, the three WDM models shown in the right panel of Figure~\ref{fig:transfer}, the three $T_{\mathrm{kd}}$ models shown in the same panel without self-interactions (hereafter ``$T_{\mathrm{kd}}$--only'' simulations), the same three $T_{\mathrm{kd}}$ models with self-interactions (hereafter ``WSIDM'' simulations), and CDM ICs with our self-interaction model (hereafter our ``SIDM'' simulation). We have resimulated all cases at lower resolution (i.e., with one fewer refinement level) and present a convergence test in Appendix~\ref{sec:convergence}. Note that our CDM, $m_{\mathrm{WDM}}=6.5~\mathrm{keV}$, and $m_{\mathrm{WDM}}=10~\mathrm{keV}$ simulations are the same as those presented in \citetalias{Nadler241003635}; our SIDM, $m_{\mathrm{WDM}}=3.5~\mathrm{keV}$ and $T_{\mathrm{kd}}$--only and WSIDM simulations are new. Thus, we present a total of eight new high-resolution simulations in this work. Our analysis is mainly based on the CDM, SIDM, $T_{\mathrm{kd}}$--only, and WSIDM simulations; we use the WDM simulations to isolate the effects of $P(k)$ suppression and test the impact of WSIDM DAOs in  Appendix~\ref{sec:wdm_comparison}. We leave a detailed comparison between our WDM and other simulations to future work.

\begin{deluxetable*}{{cccccccc}}[t!]
\centering
\tablecolumns{6}
\tablecaption{Summary of COZMIC III Simulations}
\tablehead{\colhead{Model} & \colhead{$f_{\mathrm{iso,7}}$} &
\colhead{$f_{\mathrm{iso,8}}$} &
\colhead{$f_{\mathrm{sub,7}}$} & 
\colhead{$f_{\mathrm{sub,8}}$} &
\colhead{$f_{\mathrm{cc,iso,8}}$} & \colhead{$f_{\mathrm{cc,sub,8}}$} & \colhead{Color and Linestyle}}
\startdata 
\hline \hline
CDM &
1.0 &
1.0 &
1.0 &
1.0 &
-- &
-- &
\begin{tikzpicture}[yscale=0.5] \draw [line width=0.45mm,dotted,black] (0,-1) -- (1,-1) node[right]{};; \end{tikzpicture}
\\
\hline \hline
SIDM &
0.98 & 
0.99 &
0.83 &
0.97 &
0.13 &
0.19 &
\begin{tikzpicture}[yscale=0.5] \draw [line width=0.25mm,sidm] (0,-1) -- (1,-1) node[right]{};; \end{tikzpicture} \\ 
\hline \hline
$m_{\mathrm{WDM}}=3.5~\mathrm{keV}$ & 
0.16 &
0.42 & 
0.22 &
0.49 & 
-- &
-- &
\begin{tikzpicture}[yscale=0.5] \draw [line width=0.25mm,dashed,wdm_3] (0,-1) -- (1,-1) node[right]{};; \end{tikzpicture}
\\
$m_{\mathrm{WDM}}=6.5~\mathrm{keV}$ & 
0.46 & 
0.81 & 
0.50 & 
0.81 & 
-- &
-- &
\begin{tikzpicture}[yscale=0.5] \draw [line width=0.25mm,dashed,wdm_6_5] (0,-1) -- (1,-1) node[right]{};; \end{tikzpicture}
\\
$m_{\mathrm{WDM}}=10~\mathrm{keV}$ & 
0.73 &
0.93 &
0.76 &
0.98 &
-- &
-- &
\begin{tikzpicture}[yscale=0.5] \draw [line width=0.25mm,dashed,wdm_10] (0,-1) -- (1,-1) node[right]{};; \end{tikzpicture}
\\
\hline \hline
$T_{\mathrm{kd}}=0.72~\mathrm{keV}$ & 
0.30 &
0.51 & 
0.30 &
0.48 & 
-- &
-- &
\begin{tikzpicture}[yscale=0.5] \draw [line width=0.25mm,dashed,Tkd_072] (0,-1) -- (1,-1) node[right]{};; \end{tikzpicture}
\\
$T_{\mathrm{kd}}=1.46~\mathrm{keV}$ & 
0.56 &
0.85 & 
0.61 &
0.92 & 
-- &
-- &
\begin{tikzpicture}[yscale=0.5] \draw [line width=0.25mm,dashed,Tkd_146] (0,-1) -- (1,-1) node[right]{};; \end{tikzpicture}
\\
$T_{\mathrm{kd}}=2.32~\mathrm{keV}$ & 
0.78 &
0.97 &
0.82 &
0.99 &
-- &
-- &
\begin{tikzpicture}[yscale=0.5] \draw [line width=0.25mm,dashed,Tkd_232] (0,-1) -- (1,-1) node[right]{};; \end{tikzpicture}
\\
\hline \hline
$T_{\mathrm{kd}}=0.72~\mathrm{keV}$ + SIDM & 
0.30 &
0.51 & 
0.25 &
0.46 & 
0.02 &
0.02 &
\begin{tikzpicture}[yscale=0.5] \draw [line width=0.25mm,Tkd_072] (0,-1) -- (1,-1) node[right]{};; \end{tikzpicture}
\\
$T_{\mathrm{kd}}=1.46~\mathrm{keV}$ + SIDM & 
0.55 &
0.84 & 
0.49 &
0.89 & 
0.02 &
0.08 &
\begin{tikzpicture}[yscale=0.5] \draw [line width=0.25mm,Tkd_146] (0,-1) -- (1,-1) node[right]{};; \end{tikzpicture}
\\
$T_{\mathrm{kd}}=2.32~\mathrm{keV}$ + SIDM & 
0.77 &
0.96 &
0.72 &
0.96 &
0.04 &
0.17 &
\begin{tikzpicture}[yscale=0.5] \draw [line width=0.25mm,Tkd_232] (0,-1) -- (1,-1) node[right]{};; \end{tikzpicture}
\\
\hline \hline
\enddata
{\footnotesize \tablecomments{The first column lists the DM model. The second column lists the fraction of isolated halos with $M_{\mathrm{vir}}>1.5\times 10^7~M_{\mathrm{\odot}}$ ($M_{\mathrm{vir}}>10^8~M_{\mathrm{\odot}}$) relative to CDM, denoted $f_{\mathrm{iso,7}}$ ($f_{\mathrm{iso,8}}$). Isolated halos must reside within $3~\mathrm{Mpc}$ of the host center and cannot be within the virial radius of any large halo. The third column lists the fraction of our MW host's subhalos above the same mass cuts, denoted $f_{\mathrm{sub,7}}$ ($f_{\mathrm{sub,8}}$). The fourth (fifth) column lists the fraction of core-collapsed isolated halos (subhalos) with $M_{\mathrm{vir}}>10^8~M_{\mathrm{\odot}}$, denoted $f_{\mathrm{cc,iso,8}}$ ($f_{\mathrm{cc,sub,8}}$). The core-collapsed fraction is determined using the parametric SIDM model from \cite{Yang230516176,Yang240610753}; see Section~\ref{sec:analysis}. Note that there are $583$ ($100$) subhalos and $3872$ ($759$) isolated halos with $M_{\mathrm{vir}}>1.5\times 10^7~M_{\mathrm{\odot}}$ ($M_{\mathrm{vir}}>10^8~M_{\mathrm{\odot}}$) in our CDM simulation.}}
\label{tab:summary}
\end{deluxetable*}

\begin{figure*}[t!]
\centering
\vspace{0.5cm}
\includegraphics[trim={0 11cm 0 0cm},width=0.8\textwidth]{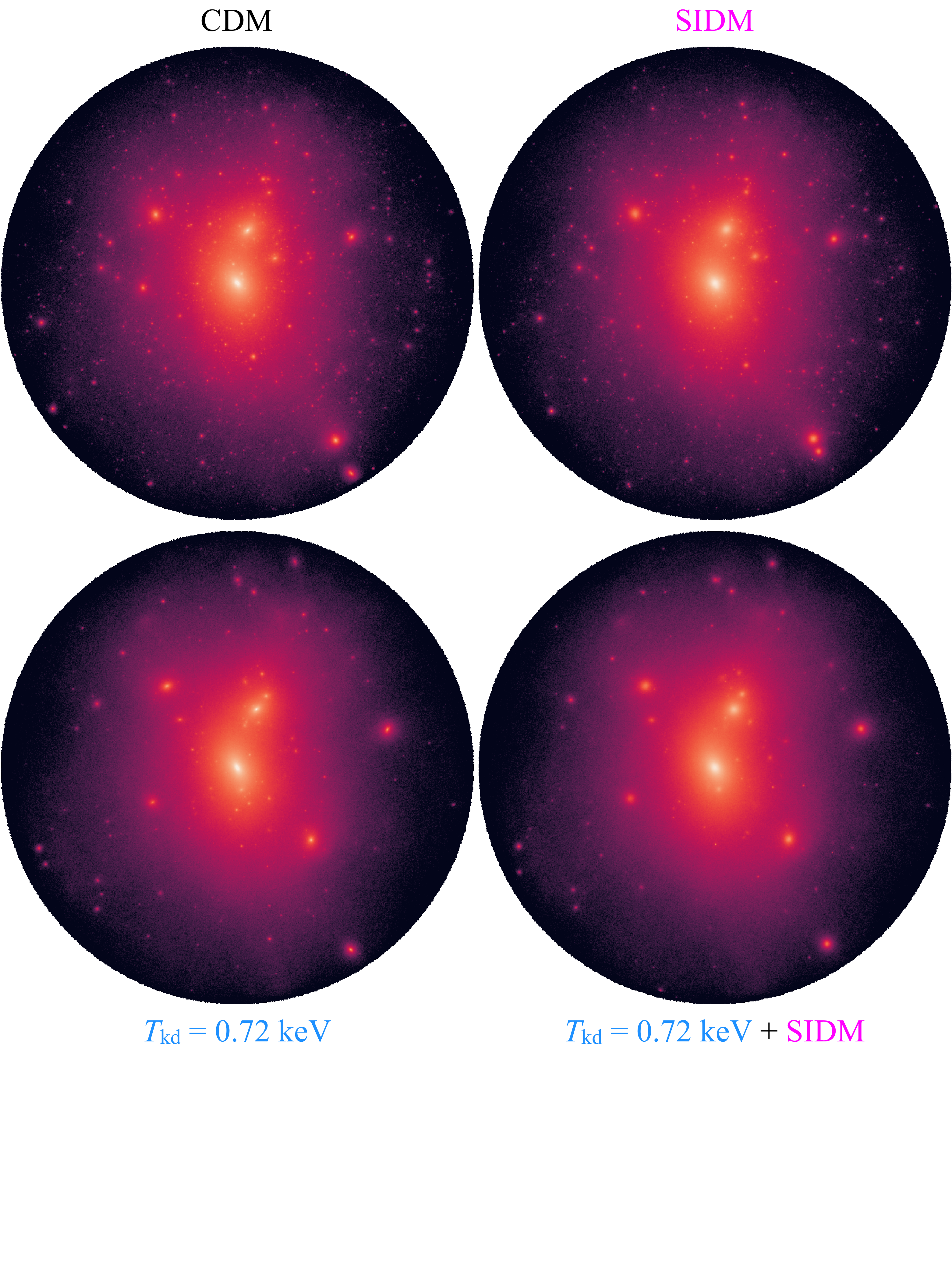}
    \caption{Projected DM density maps for a subset of our simulations: CDM (top left), SIDM only (top right), $T_{\mathrm{kd}}=0.72~\mathrm{keV}$ $P(k)$ suppression only (bottom left), and $T_{\mathrm{kd}}=0.72~\mathrm{keV}$ WSIDM (bottom right). Each visualization is centered on the host halo and spans $1.5$ times its virial radius. Visualizations were created using \textsc{meshoid} (\url{https://github.com/mikegrudic/meshoid}).}
    \label{fig:vis_main}
\end{figure*}

\subsection{Analysis Procedures}
\label{sec:analysis}

We restrict our (sub)halo abundance measurements to objects with more than $300$ particles at $z=0$, i.e., with present-day virial masses $M_{\mathrm{vir}}>1.5\times 10^7~M_{\mathrm{\odot}}$ \citep{Nadler220902675}. When analyzing (sub)halo structural properties and density profiles, we only consider objects with more than $2000$ particles at $z=0$, or $M_{\mathrm{vir}}>10^8~M_{\mathrm{\odot}}$, following \cite{Yang221113768}. This roughly selects (sub)halos with present-day maximum circular velocities $V_{\mathrm{max}}\gtrsim 7~\mathrm{km\ s}^{-1}$; we define $R_{\mathrm{max}}$ as the radius within each subhalo at which $V_{\mathrm{max}}$ is achieved. Based on \citetalias{Nadler241003635}, spurious (sub)halos formed through artificial fragmentation contribute negligibly to our WSIDM (sub)halo populations above the present-day mass resolution limit.

Throughout, we use ``peak'' to denote the largest value of a quantity achieved along a given (sub)halo's main branch, e.g., $M_{\mathrm{peak}}$ is the largest value of $M_{\mathrm{vir}}$ that a (sub)halo's most massive progenitor achieves at any snapshot. Following \citetalias{Nadler241003635}, we measure (S)HMFs using $M_{\mathrm{peak}}$ because it more closely relates to the wavenumber that sources halos of a given mass than present-day (stripped) mass, particularly for subhalos. Because $M_{\mathrm{peak}}$ functions converge slowly in the absence of a present-day mass cut (e.g., \citealt{Nadler220902675,Mansfield230810926}), we always apply the $M_{\mathrm{vir}}(z=0)$ cuts described above before measuring (S)HMFs. Note that the WDM SHMF suppression models we compare to, from \citetalias{Nadler241003635}, were derived using a similar set of cuts.

To model SIDM (sub)halos' gravothermal evolution, we apply the parametric model from \cite{Yang230516176,Yang240610753} to our simulations. This model analytically predicts SIDM density profile evolution based on CDM $V_{\mathrm{max}}$ and $R_{\mathrm{max}}$ histories. Gravothermal evolution is parameterized by $\tau\equiv t/t_c$, where $t$ is the time since a (sub)halo formed and $t_c$ is the core-collapse timescale (e.g., \citealt{Essig180901144}). Specifically, we calculate
\begin{equation}
    \tau_0 = \int_{t_f}^{t_0} \frac{\mathrm{d}t}{t_c(t)},\label{eq:tau_0}
\end{equation}
where $t_0=13.6~\mathrm{Gyr}$ is the age of the Universe today, and $t_f$ is computed for each (sub)halo using a universal halo mass--formation time relation; see \citealt{Yang230516176} for details.

We apply the model to our CDM simulation in order to predict SIDM (sub)halos' $\tau_0$ values, and to each of our $T_{\mathrm{kd}}$--only simulations to predict WSIDM (sub)halos' $\tau_0$ values. We only use (sub)halos that are matched between each pair of simulations, and we discard a handful of (sub)halos with noisy $V_{\mathrm{max}}$ and $R_{\mathrm{max}}$ histories that lead to undefined $\tau_0$ values. We set the maximum $\tau_0$ value to $1.1$, since the parametric model has been validated against SIDM simulations up to this point \citep{Yang240610753}. We define $\tau_0<0.15$ as the core-forming regime, since $\tau_0=0.15$ corresponds to the time when the lowest central density and maximum core size is achieved, and $\tau_0>0.75$ as the core-collapsed regime, since systems are rapidly collapsing at this point and have central densities that exceed corresponding (sub)halos with no self-interactions (e.g., see \citealt{Roberts240715005}).

The parametric model accurately predicts (sub)halo $V_{\mathrm{max}}$, $R_{\mathrm{max}}$, and density profile evolution histories, with an $\approx 10\%$--$50\%$ accuracy for (sub)halo density profiles when applied to cosmological CDM simulations and compared to corresponding SIDM simulations \citep{Yang240610753}. We find that the parametric model's accuracy is similar when applied to our simulations with $P(k)$ suppression and compared to our WSIDM results. Thus, we use the parametric model to compare the $\tau_0$ distribution and fraction of core-collapsed halos between our SIDM and WSIDM models. We leave detailed testing of the parametric model in WSIDM for future work.

\begin{figure*}[t!]
\centering
\includegraphics[width=0.49\textwidth]{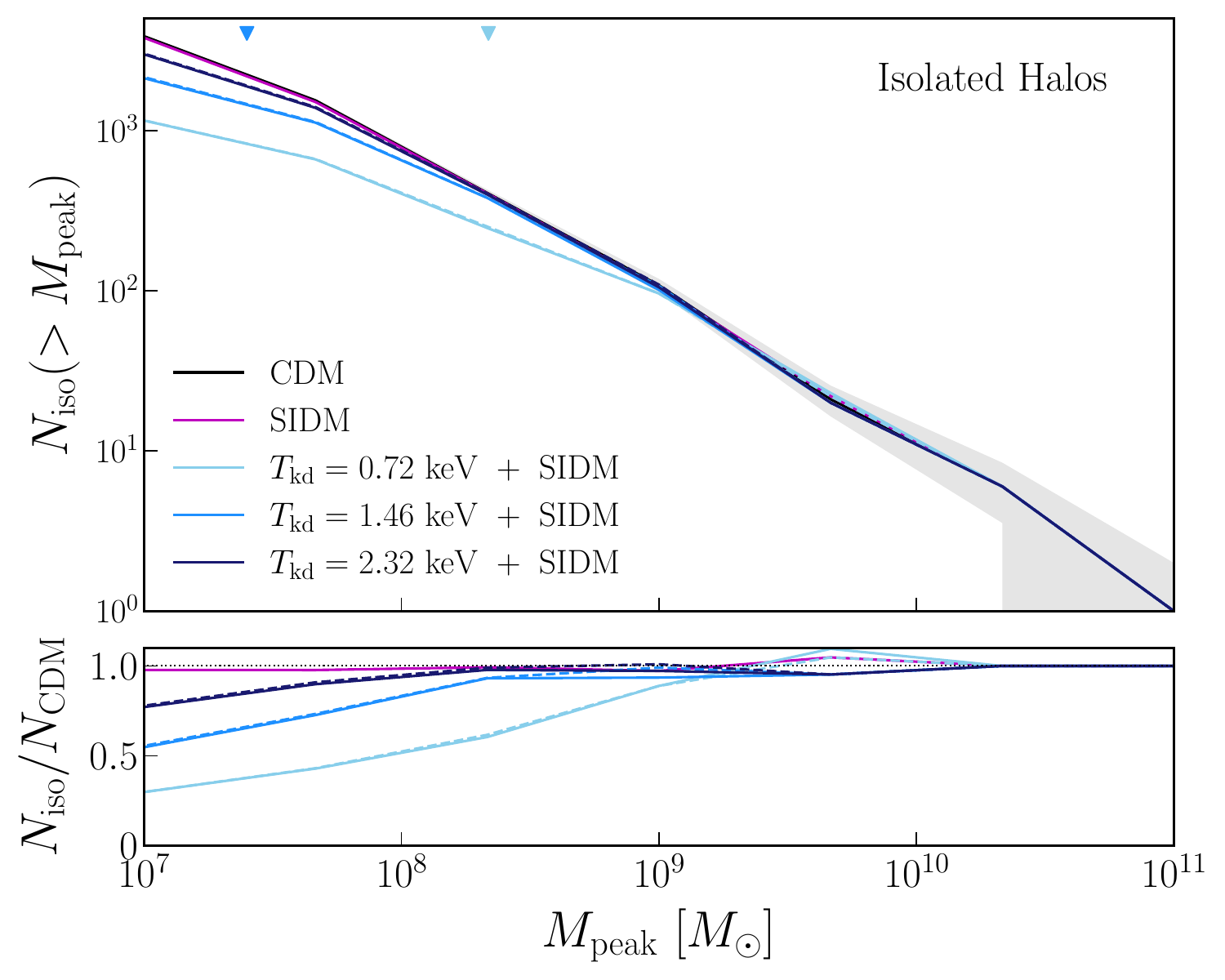}
\includegraphics[width=0.49\textwidth]{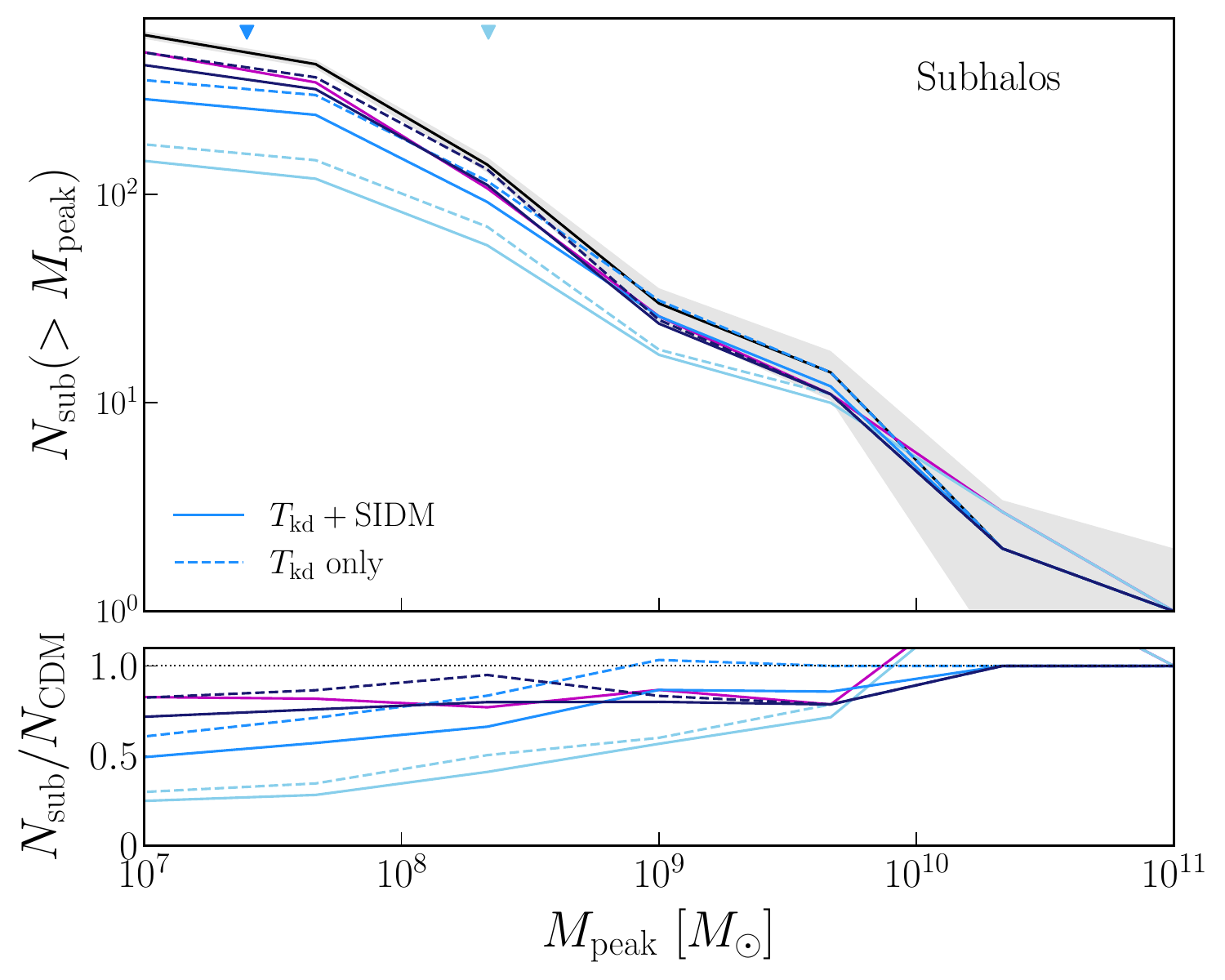}
    \caption{Cumulative isolated (left) and subhalo (right) mass functions, measured using the (sub)halo peak virial mass and subject to a cut on the present-day virial mass of $M_{\mathrm{vir}}>1.5\times 10^7~M_{\mathrm{\odot}}$. We compare our CDM simulation result (black) to SIDM (magenta), WSIDM (light to dark solid blue), and corresponding $T_{\mathrm{kd}}$--only simulations (light to dark dashed blue). Bottom panels show ratios of cumulative subhalo abundances relative to CDM. Markers indicate the half-mode masses of the $T_{\mathrm{kd}}=0.72$ and $1.46~\mathrm{keV}$ models; the half-mode mass for $T_{\mathrm{kd}}=2.32~\mathrm{keV}$ is below $10^7~M_{\mathrm{\odot}}$. Gray bands show the $1\sigma$ Poisson uncertainty on the mean cumulative CDM mass functions.}
    \label{fig:shmf}
\end{figure*}

\section{Results}
\label{sec:results}

We now present our simulation results, focusing on host halo properties (Section~\ref{sec:host}), halo and SHMF suppression relative to CDM  (Section~\ref{sec:shmf}) and structural properties, as encoded in the $R_{\mathrm{max}}$--$V_{\mathrm{max}}$ relation (Section~\ref{sec:rmax_vmax}) and core-collapsed fraction (Section~\ref{sec:tau_0}). Key results from the full set of high-resolution simulations are summarized in Table~\ref{tab:summary}, and the $z=0$ snapshots of the CDM, SIDM, $T_{\mathrm{kd}}=0.72~\mathrm{keV}$--only, and $T_{\mathrm{kd}}=0.72~\mathrm{keV}$ WSIDM simulations are visualized in Figure~\ref{fig:vis_main}.

\subsection{Host Halo Properties}
\label{sec:host}

The MW--like host's mass accretion history is nearly identical to CDM in our simulations with self-interactions and/or $P(k)$ suppression, consistent with the results of Papers~I and II for models that suppress $P(k)$. The SIDM and WSIDM simulations feature an $\mathcal{O}(\mathrm{kpc})$ core in the host's density profile, regardless of $T_{\mathrm{kd}}$; we present the corresponding host density profiles in Appendix~\ref{sec:host_density}. Note that the $P(k)$ suppression is negligible on the scale of the MW host and LMC analog, for all models we simulate (see the right panel of Figure~\ref{fig:transfer}).

\subsection{Mass Functions}
\label{sec:shmf}

In SIDM scenarios without small-scale $P(k)$ suppression, the isolated-halo mass function is not significantly suppressed relative to CDM, while the SHMF is only suppressed if the cross section is large at the subhalo infall velocity scale ($\approx 100~\mathrm{km\ s}^{-1}$, for MW subhalos; \citealt{Nadler200108754}). Thus, it is often sufficient to model SIDM effects on (sub)halo density profiles alone, without considering mass functions. Indeed, (semi)analytic models often use CDM merger trees as a template for modeling SIDM physics (e.g., \citealt{Yang230516176,Yang240610753}). However, in WSIDM, $P(k)$ suppression can significantly alter underlying (sub)halo abundances. 

To demonstrate this, Figure~\ref{fig:shmf} compares the isolated-halo (left) and subhalo (right) cumulative peak-mass functions. We find that SIDM alone does not affect isolated-halo abundances, consistent with the result from \cite{Yang221113768}. A $P(k)$ cutoff suppresses the isolated-halo abundances, and the magnitude of this suppression is not affected by the presence of self-interactions (compare the solid and dashed blue lines in the left panel of Figure~\ref{fig:shmf}); thus, isolated-halo abundances are solely determined by $P(k)$. Specifically, comparing WSIDM simulations with $T_{\mathrm{kd}}=0.72$, $1.46$, and $2.32~\mathrm{keV}$, to their CDM counterparts, we respectively find that $30\%$, $55\%$, and $77\%$ of isolated halos with $M_{\mathrm{vir}}>1.5\times 10^7~M_{\mathrm{\odot}}$ are still present; these differences are significant and larger than the statistical uncertainty on the CDM mass function.

The right panel of Figure~\ref{fig:shmf} shows that the cumulative SHMF is suppressed by $\approx 20\%$ down to $M_{\mathrm{vir}}= 1.5\times 10^7~M_{\mathrm{\odot}}$ in the SIDM simulation, consistent with \cite{Yang221113768}; also see \cite{Turner201002924} and \cite{Lovell220906834}. This suppression is due to a combination of SIDM evaporation from subhalo--host halo interactions and enhanced tidal disruption of subhalos featuring cored (rather than cuspy) profiles (e.g., \citealt{Vogelsberger12015892,Nadler200108754}). Introducing a $P(k)$ cutoff within WSIDM simulations further enhances the suppression: only $25\%$, $49\%$, and $72\%$ of subhalos with $M_{\mathrm{vir}}>1.5\times 10^7~M_{\mathrm{\odot}}$ remain, relative to CDM, in the $T_{\mathrm{kd}}=0.72$, $1.46$, and $2.32~\mathrm{keV}$ cases, respectively; again, these differences are significant, given the statistical uncertainty on the CDM SHMF.\footnote{The SIDM SHMF is slightly more suppressed than the $T_{\mathrm{kd}}=1.46$ and $2.32~\mathrm{keV}$ WSIDM SHMFs at $M_{\mathrm{peak}}\approx 10^{9.5}$ and $10^{8.5}~M_{\mathrm{\odot}}$, respectively, but these differences are not statistically significant.} Compared to our simulations with $P(k)$ suppression alone, WSIDM SHMFs are slightly more suppressed, at the $\approx 5\%$ level (compare the solid and dashed lines in the bottom right panel of Figure~\ref{fig:shmf}). Thus, SHMF suppression is mainly determined by the $P(k)$ cutoff, rather self-interactions for the models we simulate. 

Our $T_{\mathrm{kd}}=0.72~\mathrm{keV}$ WSIDM SHMF suppression is weaker than the simulation predictions for ETHOS-2 and ETHOS-3 models in \cite{Vogelsberger151205349}, which included $P(k)$ suppression and an SIDM cross section comparable to our WSIDM models. This is consistent with the slightly stronger $P(k)$ suppression in ETHOS-2 and ETHOS-3 models compared to our $T_{\mathrm{kd}}=0.72~\mathrm{keV}$ case. In Section~\ref{sec:f_cc}, we will demonstrate that core collapse is almost entirely avoided in models with such extreme $P(k)$ suppression. Thus, SHMF suppression may also be more severe in \cite{Vogelsberger151205349} compared to our results because their subhalos are uniformly cored, and thus more susceptible to tides than our subhalo populations, which contain core-collapsed systems.

We compare our $T_{\mathrm{kd}}$--only halo and SHMFs to our WDM simulations in Appendix~\ref{sec:wdm_comparison}, finding that the WDM models yield $\approx 10\%$ stronger suppression of total (sub)halo abundances compared to half-mode-matched $T_{\mathrm{kd}}$ models. This is most likely due to the different slopes of the $P(k)$ suppression in WDM versus our WSIDM models. The DAOs in our WSIDM ICs could also contribute to the difference by slightly reducing halo and subhalo abundance suppression relative to the half-mode-matched WDM models, although we expect this effect to be small based on the results in \citetalias{Nadler241003635}.

\subsection{$R_{\mathrm{max}}$--$V_{\mathrm{max}}$ Relations}
\label{sec:rmax_vmax}

\begin{figure*}[t!]
\centering
\includegraphics[width=\textwidth]{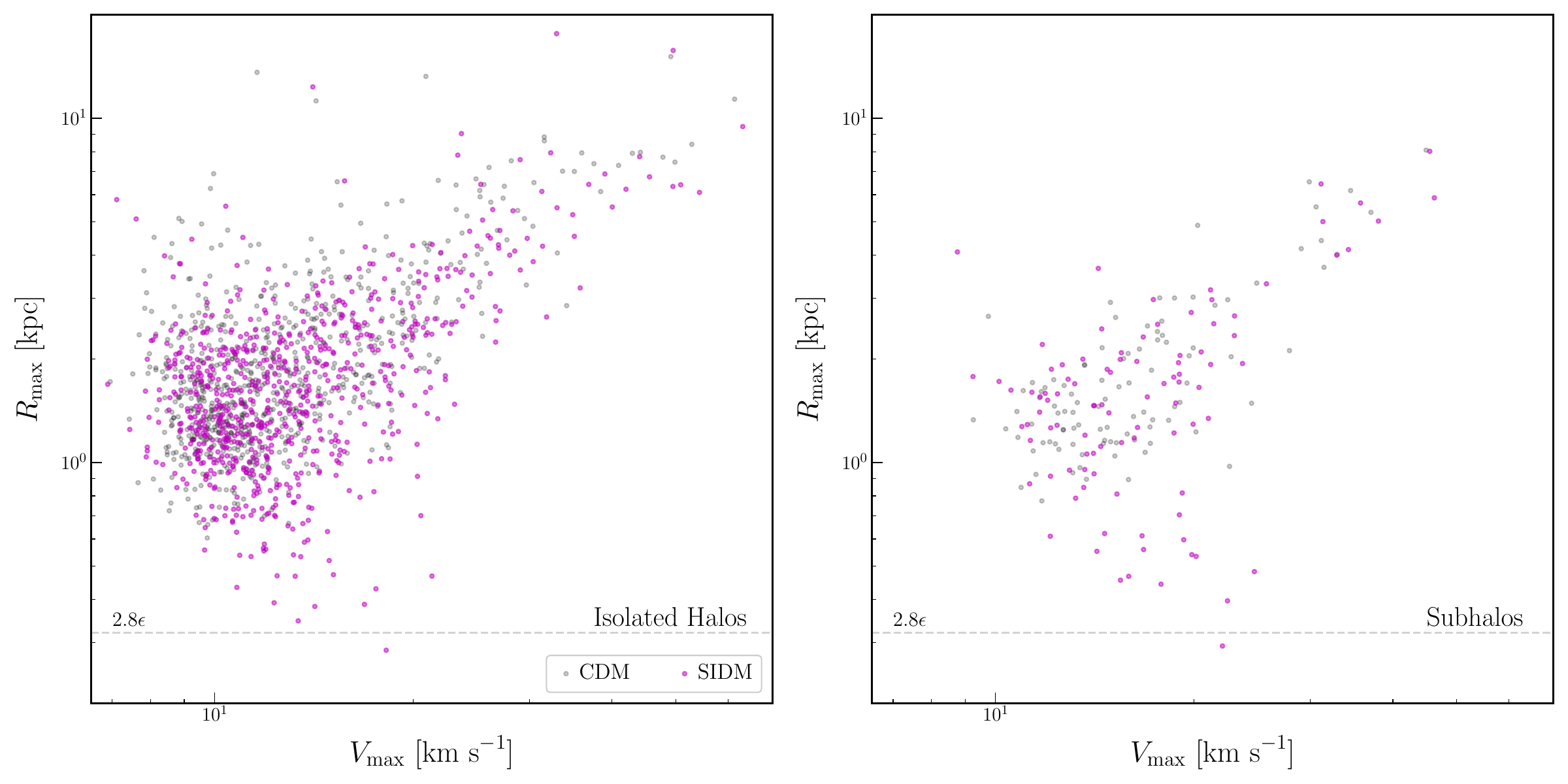}
\includegraphics[width=\textwidth]{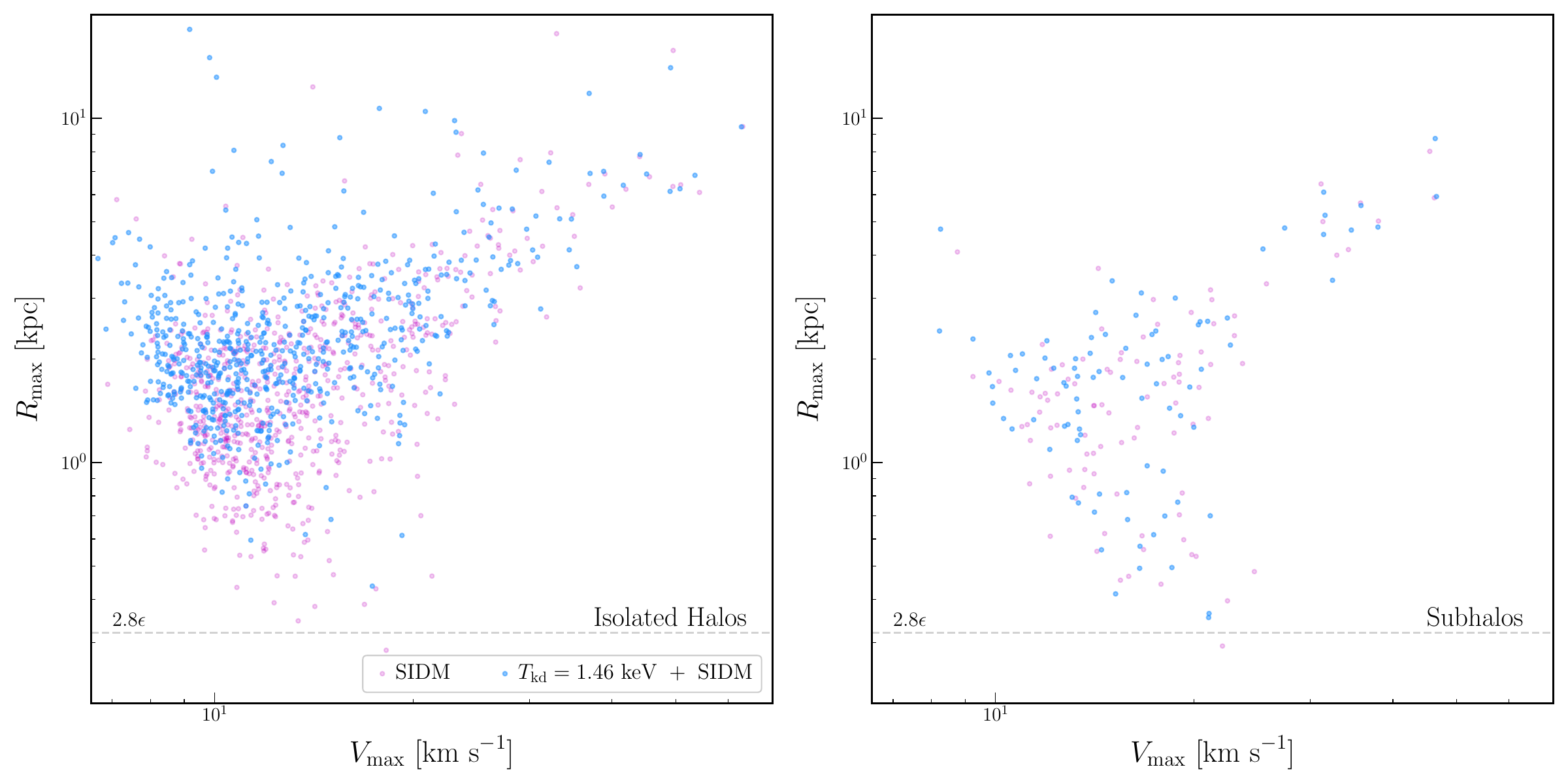}
    \caption{The relation between maximum circular velocity, $V_{\mathrm{max}}$, and the radius at which $V_{\mathrm{max}}$ is achieved, $R_{\mathrm{max}}$, for isolated halos (left column) and subhalos (right column). Results are shown for our CDM simulation (gray), our SIDM simulation with CDM ICs (magenta), and our SIDM simulation with $T_{\mathrm{kd}}=1.46~\mathrm{keV}$ ICs (blue).}
    \label{fig:tau_0_vmax}
\end{figure*}

The $R_{\mathrm{max}}$--$V_{\mathrm{max}}$ relation encodes the structural properties of halos without assuming a specific density profile, and has thus been shown to effectively encode all stages of SIDM gravothermal evolution \citep{Yang221113768,Yang230516176,Ando240316633}. Figure~\ref{fig:tau_0_vmax} shows the isolated halo and subhalo $R_{\mathrm{max}}$--$V_{\mathrm{max}}$ relation in our CDM, SIDM, and $T_{\mathrm{kd}}=1.46~\mathrm{keV}$ WSIDM simulations. The comparison between CDM and SIDM in the top panels is consistent with the findings in \cite{Yang221113768}: MilkyWaySIDM self-interactions shift a sizable fraction of low-mass isolated halos and subhalos toward larger $V_{\mathrm{max}}$ and smaller $R_{\mathrm{max}}$ ($\lesssim 1~\mathrm{kpc}$) compared to their CDM counterparts, indicating core collapse.\footnote{The smallest values of $R_{\mathrm{max}}$ we measure are comparable to the convergence radius of our simulations,  $2.8\epsilon\approx 320~\mathrm{pc}$ \citep{Ludlow2019}.} The remaining core-forming systems shift toward larger $R_{\mathrm{max}}$ than in CDM. Note that low-$R_{\mathrm{max}}$ systems form a larger fraction of the total subhalo (versus isolated halo) population because tidal stripping accelerates core collapse for the MilkyWaySIDM cross section \citep{Kahlhoefer190410539,Nishikawa190100499,Sameie190407872,Yang210202375,Zeng211000259}.

\begin{figure*}[t!]
\centering
\includegraphics[width=\textwidth]{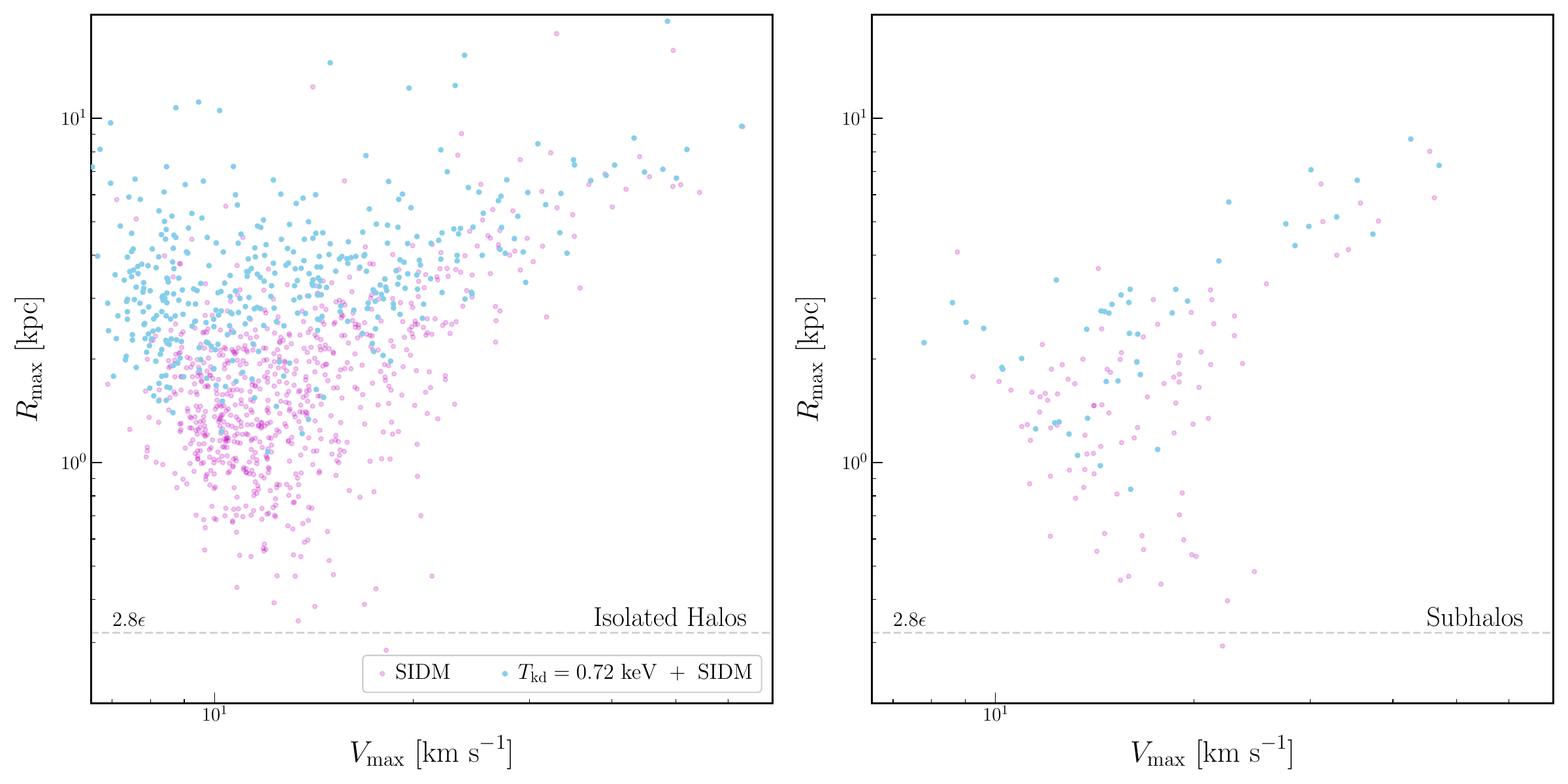}
\includegraphics[width=\textwidth]{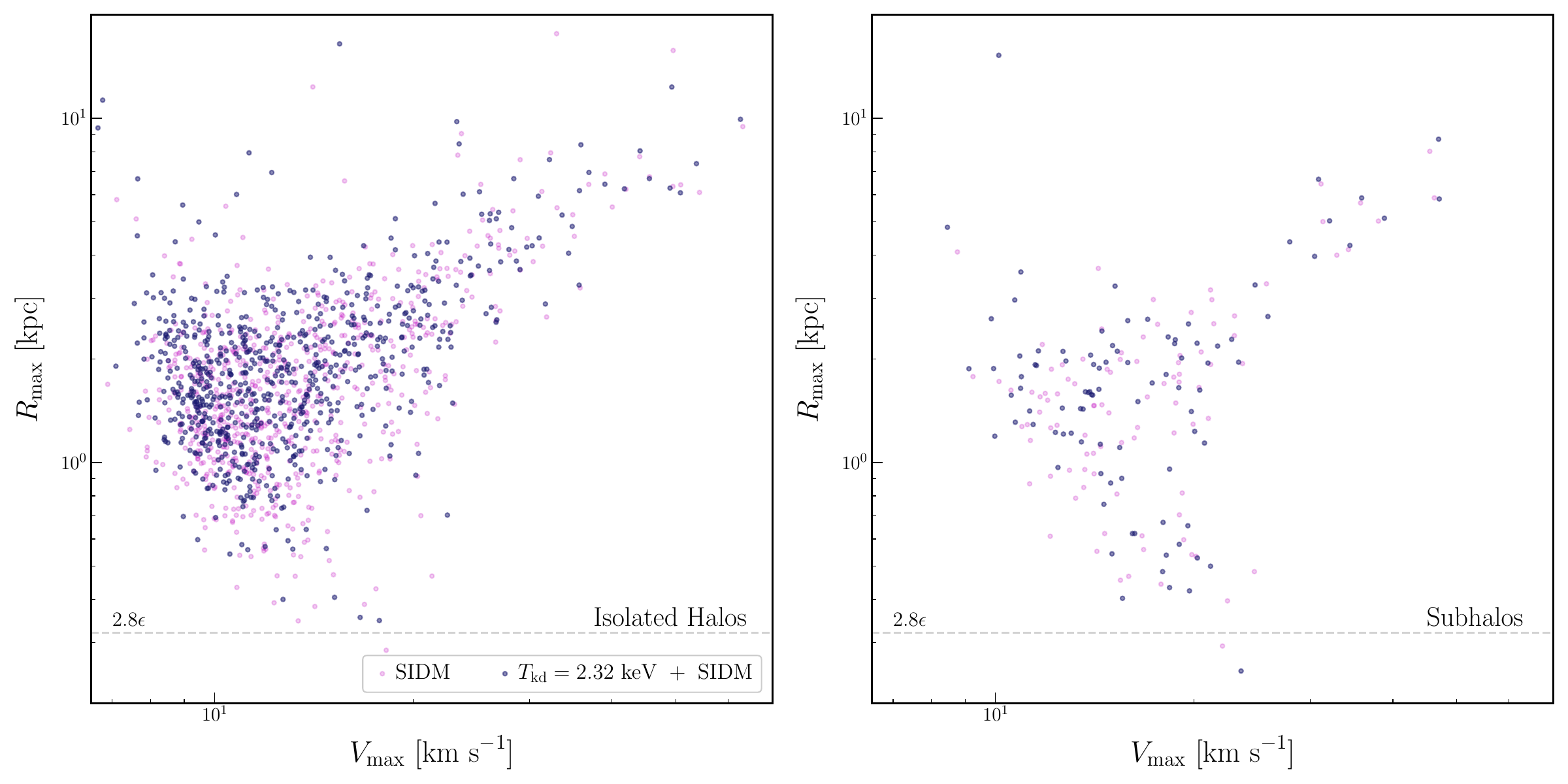}
    \caption{Same as Figure~\ref{fig:tau_0_vmax}, for our WSIDM simulation with $T_{\mathrm{kd}}=0.72~\mathrm{keV}$ (top) and $2.32~\mathrm{keV}$ (bottom); in addition, both panels show the result for our SIDM simulation (magenta).}
    \label{fig:tau_0_vmax_alt}
\end{figure*}

Introducing $P(k)$ suppression significantly affects the $R_{\mathrm{max}}$--$V_{\mathrm{max}}$ relation. For example, the bottom panels of Figure~\ref{fig:tau_0_vmax} show that (sub)halos in our $T_{\mathrm{kd}}=1.46~\mathrm{keV}$ WSIDM simulation are shifted toward larger $R_{\mathrm{max}}$ than in SIDM. A few isolated halos and a moderate number of subhalos with small values of $R_{\mathrm{max}}$ remain in this WSIDM scenario, but they form a much smaller fraction of the total populations than in SIDM. Figure~\ref{fig:tau_0_vmax_alt} shows that there is a similar shift toward larger $R_{\mathrm{max}}$ for our $T_{\mathrm{kd}}=0.72~\mathrm{keV}$ and $T_{\mathrm{kd}}=2.32~\mathrm{keV}$ WSIDM simulations. In the $T_{\mathrm{kd}}=0.72~\mathrm{keV}$ WSIDM simulation, low-$R_{\mathrm{max}}$ halos are almost entirely erased. Thus, the shift toward larger $R_{\mathrm{max}}$ is more pronounced for models with more severe $P(k)$ suppression (i.e., lower $T_{\mathrm{kd}}$). In addition, the suppression of the core-collapsed population is more significant for isolated halos versus subhalos when $P(k)$ suppression is introduced. 

Note that $P(k)$ suppression alone contributes to the shift toward larger values of $R_{\mathrm{max}}$ due to the delayed formation times and reduced concentrations of low-mass (sub)halos in the presence of a cutoff. This effect persists and can even be exacerbated when SIDM is included, since (sub)halos with delayed formation times are more likely to be found in the core-forming stage. This increased abundance of isolated, high-$R_{\mathrm{max}}$ halos may help to explain the extremely low halo concentrations of gas-rich isolated ultradiffuse galaxies \citep{ManceraPina:2019zih,ManceraPina:2020ujo,PinaMancera211200017,Kong220405981}. We leave this direction for future work.

\begin{figure*}[t!]
\centering
\includegraphics[width=\textwidth]{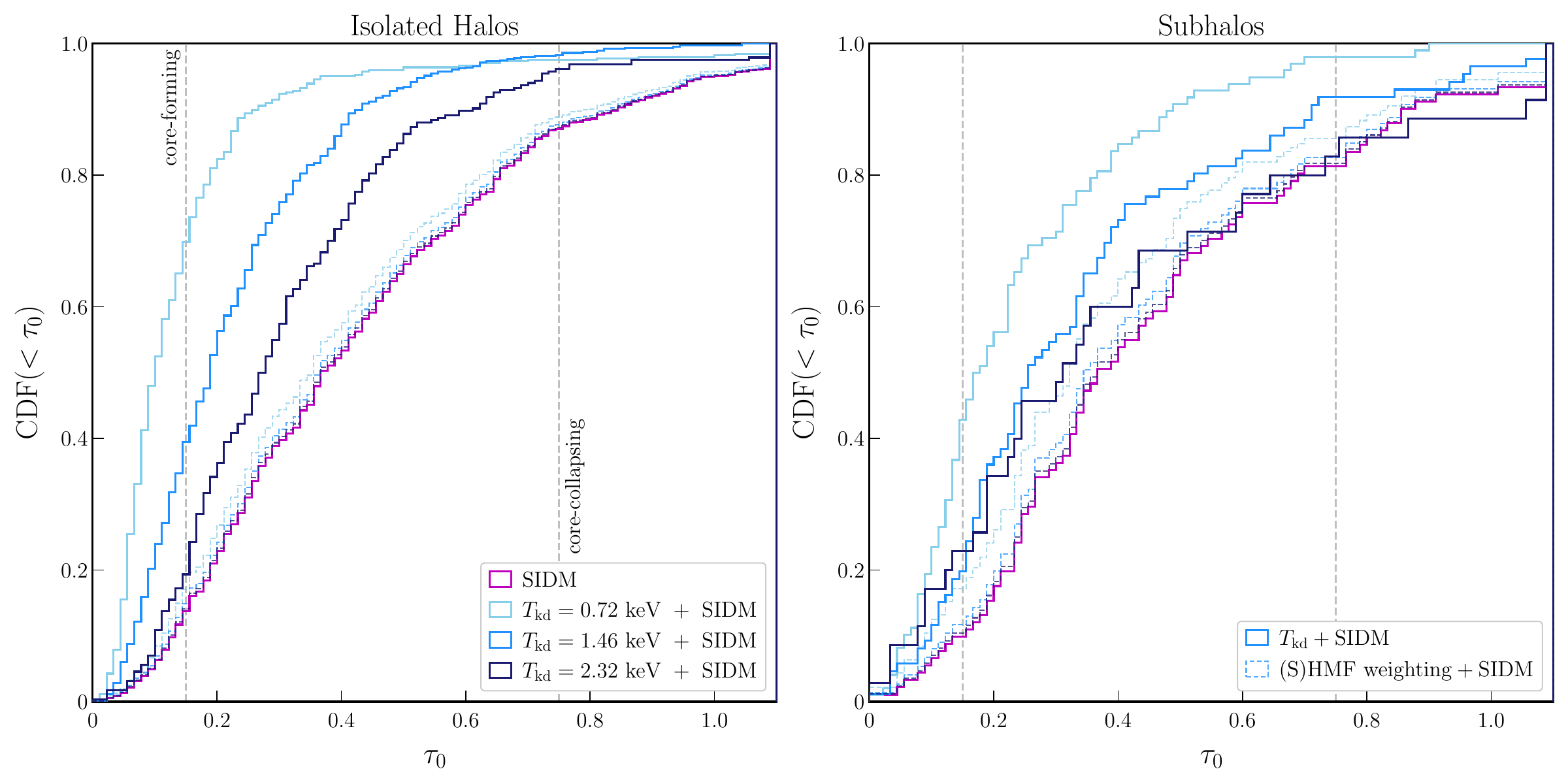}
    \caption{Normalized cumulative distributions of the gravothermal evolution timescale, $\tau_0$, for isolated halos (left) and subhalos (right) with $M_{\mathrm{vir}}>10^8~M_{\mathrm{\odot}}$ in our SIDM (magenta) and $T_{\mathrm{kd}}=0.72$, $1.46$, and $2.32~\mathrm{keV}$ WSIDM simulations (thick solid lines). The gravothermal evolution timescale $\tau_0$ is calculated using the parametric SIDM model from \cite{Yang230516176,Yang240610753}, and clipped at a maximum value of $1.1$; see Section~\ref{sec:analysis} for details. Systems with $\tau_0<0.15$ (left dashed vertical line) are in the core expansion phase; systems with $\tau_0>0.75$ (right dashed vertical line) are deep in the core-collapsed phase, with a rapidly increasing central density that exceeds its original value. In both panels, thick solid lines show the WSIDM simulation result, while thin dashed lines show the SIDM $\tau_0$ distribution weighted by the probability that (sub)halos form in each $T_{\mathrm{kd}}$ model based on an effective mapping to WDM (see Section~\ref{sec:f_cc}).}
    \label{fig:tau_0}
\end{figure*}

\subsection{Gravothermal Evolution Timescales}
\label{sec:tau_0}

We now use the parametric model to predict gravothermal evolution timescales following the procedure described in Section~\ref{sec:analysis}. For both isolated halos and subhalos, we apply an $M_{\mathrm{vir}}>10^8~M_{\mathrm{\odot}}$ cut to ensure that their profiles are well resolved. Figure~\ref{fig:tau_0} shows the normalized cumulative $\tau_0$ distribution for isolated halos (left) and subhalos (right). As $P(k)$ suppression is introduced and becomes more severe, the fraction of core-forming isolated halos increases approximately linearly across our models. The subhalo $\tau_0$ distribution is affected in a less regular fashion, particularly when comparing our SIDM, and $T_{\mathrm{kd}}=1.46$ and $2.32~\mathrm{keV}$ WSIDM results. This is consistent with the difference in $R_{\mathrm{max}}$--$V_{\mathrm{max}}$ relations we observed for subhalos versus isolated halos in Section~\ref{sec:rmax_vmax}, and further confirms that linear $P(k)$ suppression and nonlinear subhalo evolution physics affect (sub)halos' gravothermal evolution in a nontrivial manner; we study these coupled effects further in Section~\ref{sec:f_cc}.

The fraction of core-collapsed subhalos (isolated halos) with $M_{\mathrm{vir}}>10^8~M_{\mathrm{\odot}}$ in our SIDM simulation is $18\%$ ($13\%$), consistent with the results in \cite{Yang221113768}. The core-collapsed fraction is expected to peak at roughly $10^8~M_{\mathrm{\odot}}$, which corresponds to the velocity scale $w$ of the turnover in the underlying SIDM cross section \citep{Ando240316633}. In our $T_{\mathrm{kd}}=0.72$, $1.46$, and $2.32~\mathrm{keV}$ models, we predict that these subhalo (isolated halo) core-collapsed fractions become $2\%$ ($2\%$), $8\%$ ($2\%$), and $17\%$ ($4\%$), respectively. 

For isolated halos, these changes to the core-collapsed fraction primarily probe suppression in halo growth histories (and thus concentrations) due to $P(k)$. Since $\tau_0$ depends strongly on halo concentration, even the $T_{\mathrm{kd}}=2.32~\mathrm{keV}$ suppression is sufficient to delay core collapse in nearly all isolated halos we resolve. Meanwhile, the core-collapsed fraction for subhalos is sensitive to the correlated effects of growth history and tidal stripping. Since core-collapsed subhalos are more resistant to tidal disruption than cored systems, the subhalo core-collapsed fraction is less sensitive to $P(k)$ suppression for the coldest models we simulate.

\section{The Interplay between $P(k)$ Suppression and Gravothermal Evolution}
\label{sec:f_cc}

\begin{figure*}[t!]
\hspace{-25mm}
\includegraphics[width=1.25\textwidth]{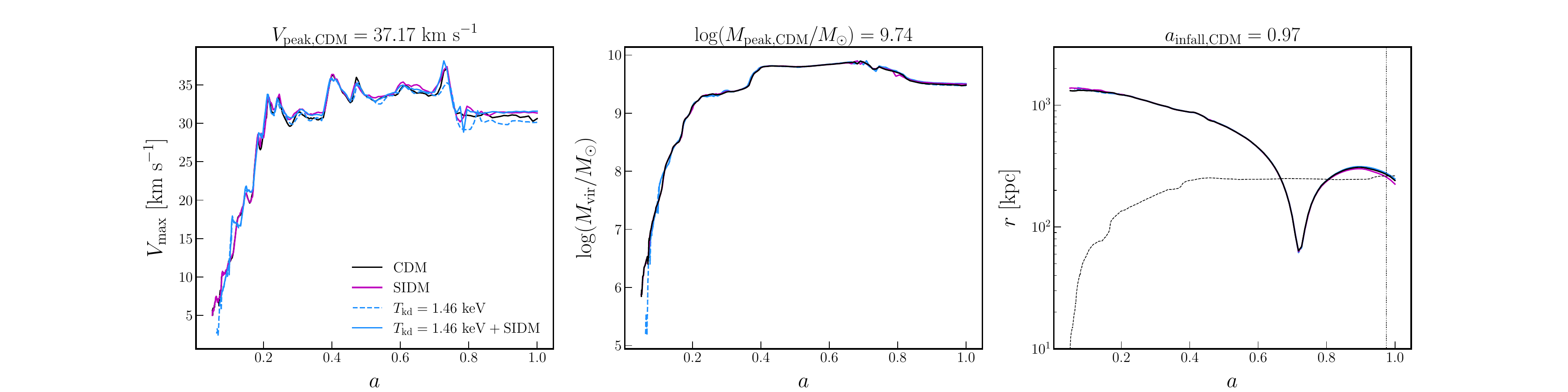} \\
\hspace*{-25mm}
\includegraphics[width=1.25\textwidth]{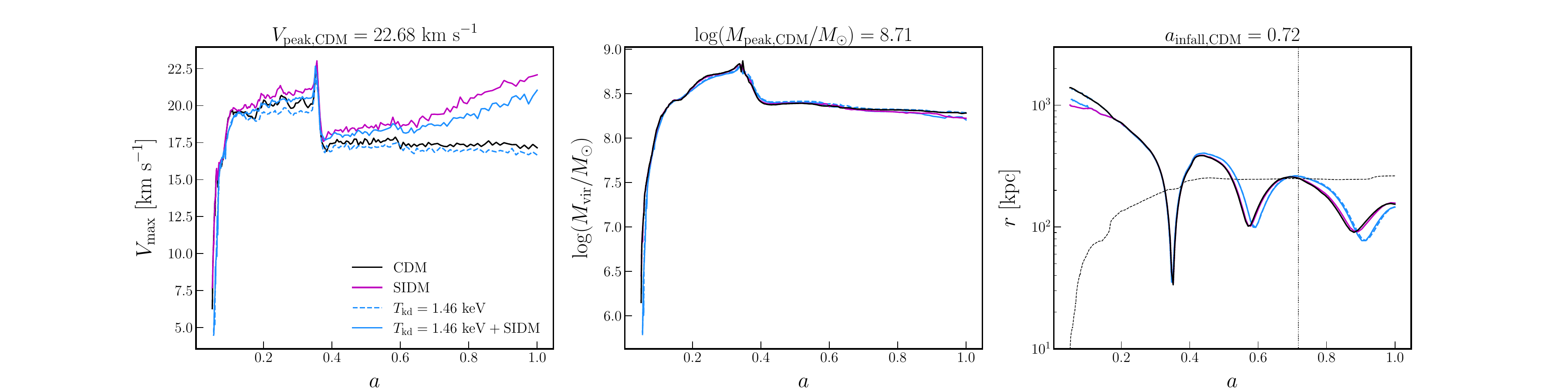} \\
\hspace*{-25mm}
\includegraphics[width=1.25\textwidth]{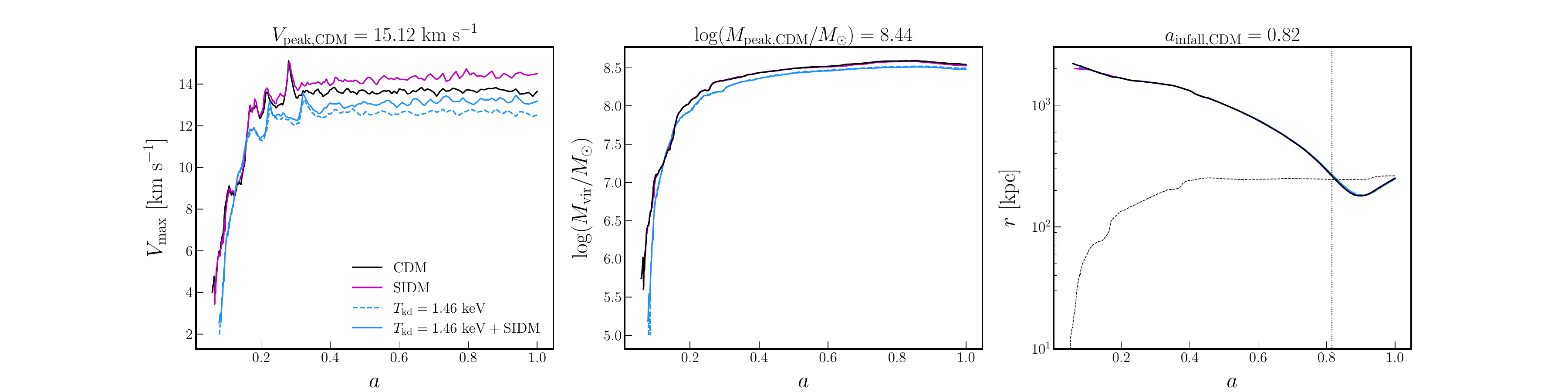}
    \caption{Evolution of maximum circular velocity $V_{\mathrm{max}}$ (left), virial mass $M_{\mathrm{vir}}$ (middle), and distance from the host center $r$ (right) vs.\ scale factor $a$ for matched subhalos. Results for a high (top), medium (middle), and low-mass (bottom) subhalo are shown for CDM (black), SIDM (magenta), $T_{\mathrm{kd}}=1.46~\mathrm{keV}$ WSIDM (blue), and $T_{\mathrm{kd}}=1.46~\mathrm{keV}$--only (dashed blue). In the right column, the dashed black line shows the MW host's virial radius, and the vertical dotted line shows the scale factor of each subhalo's most recent infall into the host, $a_{\mathrm{infall}}$.}
    \label{fig:matched_halos}
\end{figure*}

We now study the physical drivers of WSIDM (sub)halos' gravothermal evolution. When comparing our WSIDM and SIDM simulations, two main effects are at play:
\begin{enumerate}
    \item some low-mass (sub)halos that core collapse in SIDM (without $P(k)$ suppression) do not form in WSIDM;
    \item some (sub)halos that core collapse in SIDM have delayed and/or suppressed growth histories in WSIDM, preventing core collapse.
\end{enumerate}

To assess the impact of the first effect (i.e., the erasure of low-mass (sub)halos), we fit ``effective'' WDM models that match the total halo and subhalo abundances down to the $M_{\mathrm{vir}}=1.5\times 10^7~M_{\mathrm{\odot}}$ resolution limit in our $T_{\mathrm{kd}}$--only simulations, using the WDM SHMF suppression fit derived in \citetalias{Nadler241003635} (see Appendix~\ref{sec:wdm_comparison} for details). For simplicity, we assume this fit also applies to the isolated-halo mass function suppression. We derive effective thermal-relic WDM masses of $m_{\mathrm{WDM}}=4.2$ ($5.3$), $6.9$ ($7.9$), and $10.2$ ($11.3$) $\mathrm{keV}$ for subhalos (isolated halos) in our $T_{\mathrm{kd}}=0.72$ and $1.46$, and $2.32~\mathrm{keV}$ simulations, respectively. These values are slightly larger than the masses of the corresponding half-mode-matched WDM models due to differences in the shape of the transfer function cutoff and/or the low-amplitude DAOs in our $T_{\mathrm{kd}}$ models, as discussed in Section~\ref{sec:shmf}.

We apply this suppression to reweigh the predicted $\tau_0$ distribution from our SIDM simulation, based on the (sub)halos' $M_{\mathrm{peak}}$ values. These results, shown by the faint dashed lines in Figure~\ref{fig:tau_0}, are nearly identical to the SIDM (magenta) result for isolated halos. Thus, the erasure of low-mass systems has a minor effect on the $\tau_0$ distributions, consistent with our conclusion above that the isolated-halo core-collapsed fraction is primarily sensitive to halo growth histories. For subhalos, the erasure of low-mass systems explains about half of the shift in the $T_{\mathrm{kd}}=0.72~\mathrm{keV}$ WSIDM $\tau_0$ distribution relative to SIDM; the shift is smaller for $T_{\mathrm{kd}}=1.46$ and $2.32~\mathrm{keV}$. 

To assess the impact of the second effect (delayed and suppressed growth), we identify pairs of matched subhalos between our simulations by minimizing the cumulative difference between their orbital and $V_{\mathrm{max}}$ histories since the time each subhalo first becomes well resolved; we have confirmed that this matching procedure robustly identifies objects that contain the same bound particles when ICs are fixed (e.g., when comparing CDM and SIDM). Figure~\ref{fig:matched_halos} illustrates the evolution of three representative pairs of matched subhalos from our CDM, SIDM, and $T_{\mathrm{kd}}=1.46~\mathrm{keV}$ simulations with and without self-interactions.
\begin{itemize}
    \item The high-mass subhalo (top row) has a slightly delayed merger history and suppressed $V_{\mathrm{max}}$ evolution in our $T_{\mathrm{kd}}$--only simulation, relative to CDM. In SIDM, its final value of $V_{\mathrm{max}}$ is slightly enhanced relative to CDM and similar to the WSIDM result.
    \item The moderate-mass subhalo (middle row) is core collapsed in our SIDM simulation, signaled by its steeply rising $V_{\mathrm{max}}$ evolution at late times relative to CDM. This subhalo is less collapsed in our WSIDM simulation, while its $V_{\mathrm{max}}$ history is suppressed in our $T_{\mathrm{kd}}$--only simulation. Mass-loss rates in our SIDM and WSIDM simulations are slightly enhanced relative to CDM. In addition, its orbital phase shifts in the $T_{\mathrm{kd}}$ simulations with and without self-interactions. 
    \item The low-mass subhalo (bottom row) has a significantly suppressed (and mildly delayed) $V_{\mathrm{max}}$ evolution history in the $T_{\mathrm{kd}}=1.46~\mathrm{keV}$--only WSIDM run relative to CDM. In SIDM, this subhalo is core collapsed (evidenced by its enhanced $V_{\mathrm{max}}$ relative to CDM). In WSIDM, its $V_{\mathrm{max}}$ is suppressed at early times but nearly catches up to CDM at late times, while the $T_{\mathrm{kd}}$--only version retains a lower $V_{\mathrm{max}}$ throughout its history.
    \end{itemize}

Thus, introducing $P(k)$ suppression can delay or even erase core collapse. In our $T_{\mathrm{kd}}=1.46~\mathrm{keV}$ WSIDM simulation, this is mainly due to the reduced concentration of WSIDM (sub)halos, which systematically achieve lower values of $V_{\mathrm{peak}}$ than in CDM. Crucially, the (sub)halos' early evolution histories significantly affect their subsequent gravothermal evolution, as the core-collapse timescale is extremely sensitive to halo concentration \citep{Essig180901144,Nadler230601830}. As a result, two competing effects---core collapse driven by SIDM and suppressed growth driven by $P(k)$---can roughly cancel for some (sub)halos, resulting in present-day $V_{\mathrm{max}}$ values similar to CDM.

We also show matched subhalo evolution histories for $T_{\mathrm{kd}}=0.72$ and $2.32~\mathrm{keV}$ in Appendix~\ref{sec:matched_alt}. In the $T_{\mathrm{kd}}=0.72~\mathrm{keV}$ case, the WSIDM subhalo evolution histories are substantially suppressed and delayed relative to SIDM. Thus, in addition to the effects of reduced concentration discussed above, gravothermal evolution in models with extreme $P(k)$ suppression has less time to operate, further reducing the core-collapsed fraction. Meanwhile, in the $T_{\mathrm{kd}}=2.32~\mathrm{keV}$ case, the delay is negligible and the reduced halo concentration effect dominates over delayed growth. Beyond the examples shown in Figure~\ref{fig:matched_halos} and Appendix~\ref{sec:matched_alt}, we have verified that these conclusions are robust by inspecting the $V_{\mathrm{max}}$ histories of all matched subhalos above our resolution limit. In future work, it will be important to study how the effects discussed above propagate to (sub)halos' full density profiles.

Finally, we note that the tidal evolution of WSIDM subhalos can differ from that of matched SIDM subhalos; in turn, this can alter their gravothermal evolution. In particular, tidal evolution can be affected by changes to the host potential and/or subhalo density profiles between the SIDM and WSIDM models. In Appendix~\ref{sec:host_density}, we show that the MW host's density profile is nearly identical in our SIDM and WSIDM simulations, indicating that the latter effect is more important given our SIDM cross section.

\section{Predictions for Dwarf Galaxy Subhalos}
\label{sec:dwarf}

We now study the density profiles (Section~\ref{sec:dwarf_dens}) and central density--pericenter relation (Section~\ref{sec:dwarf_peri}) of subhalos that are expected to host dwarf galaxies in our simulations. Throughout this section, we study the $T_{\mathrm{kd}}=1.46~\mathrm{keV}$ model because it is closest to the current WDM constraint from MW satellite galaxy population derived by \cite{Nadler200800022}; we show results for other $T_{\mathrm{kd}}$ models in Appendix~\ref{sec:dwarf_alt}. We focus on subhalos, with an eye toward modeling MW satellite galaxies.

We use the abundance-matching model from \cite{Nadler180905542,Nadler191203303,Nadler240110318}, with best-fit parameters inferred from Dark Energy Survey and Pan-STARRS1 MW satellite observations \citep{Drlica-Wagner191203302,Nadler191203303}, to predict the luminosity of the satellite galaxy hosted by each subhalo as a function of its peak maximum circular velocity, $V_{\mathrm{peak}}$. We measure $V_{\mathrm{peak}}$ using only snapshots prior to infall, rather than all snapshots, to avoid increases in the predicted luminosity due to late-time increases in $V_{\mathrm{max}}$ for core-collapsed SIDM and WSIDM subhalos. Following \cite{Nadler240110318}, we do not alter the galaxy--halo connection model in our beyond-CDM simulations; \cite{Despali250112439} show that assumption is reasonable for higher-mass WDM and SIDM systems than we consider. We leave a study of the coupling between $P(k)$ suppression, SIDM effects, and the galaxy--halo connection to future work.

\subsection{Subhalo Density Profiles}
\label{sec:dwarf_dens}

\begin{figure*}[t!]
\centering
\includegraphics[width=0.925\textwidth,trim={0 0.375cm 0 0cm}]{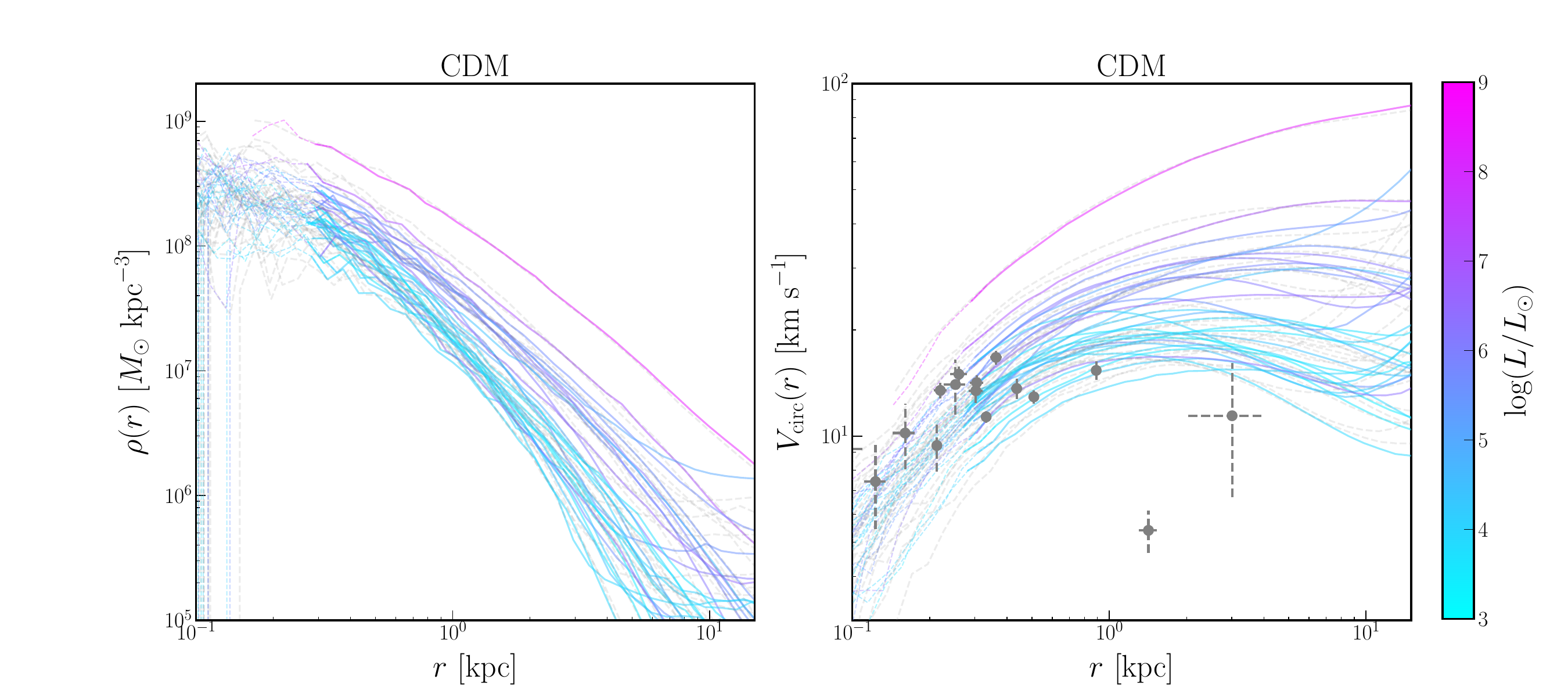} 
\includegraphics[width=0.925\textwidth,trim={0 0.375cm 0 0cm}]{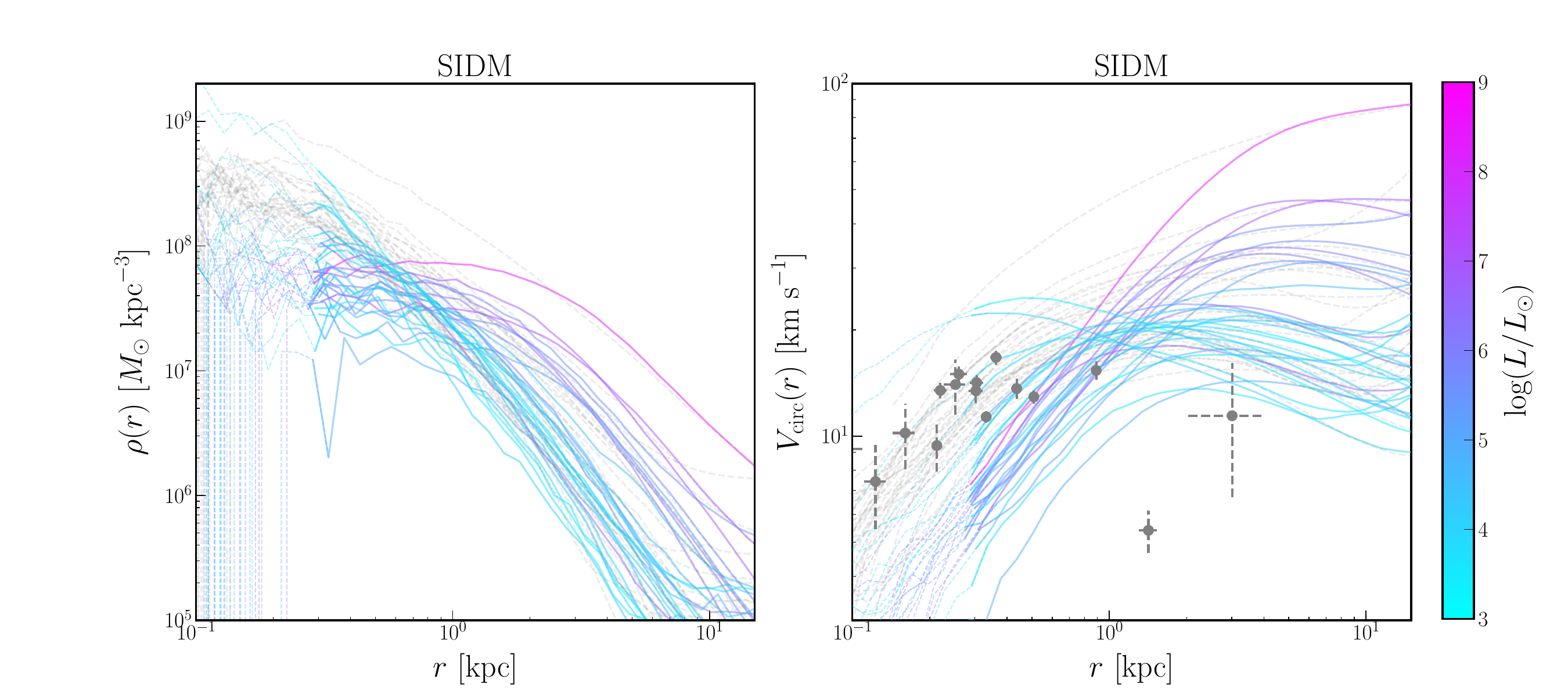} 
\includegraphics[width=0.925\textwidth,trim={0 0.375cm 0 0cm}]{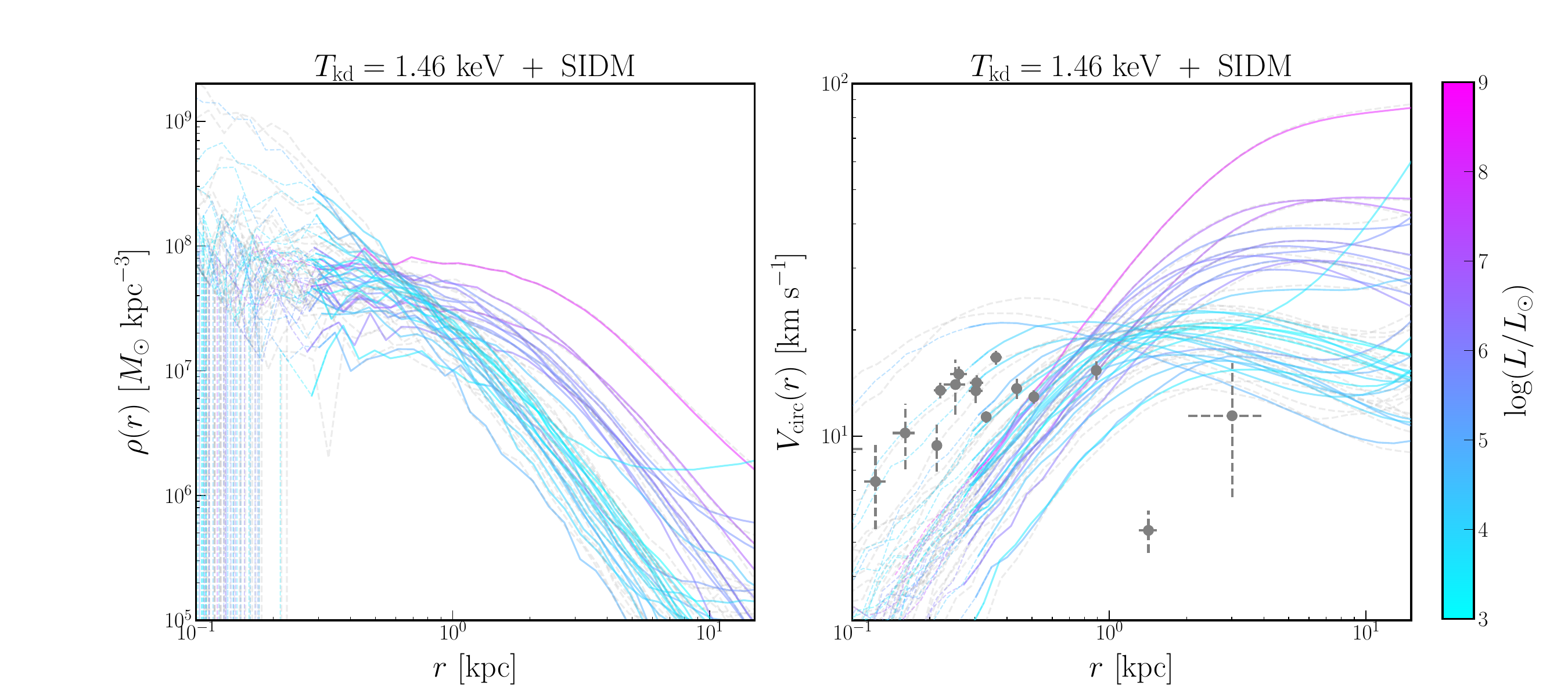}
    \caption{Density profiles (left) and circular velocity profiles (right) of the $30$ highest-$V_{\mathrm{peak}}$ subhalos with $M_{\mathrm{vir}}>1.2\times 10^8~M_{\mathrm{\odot}}$ in our CDM (top), SIDM (middle), and $T_{\mathrm{kd}}=1.46~\mathrm{keV}$ WSIDM (bottom) simulations. Faint gray lines in the top, middle, and bottom rows respectively show $T_{\mathrm{kd}}=1.46~\mathrm{keV}$--only, CDM, and SIDM results. Subhalos are colored by the luminosities of the dwarf galaxies they are predicted to host using the \cite{Nadler191203303} abundance-matching model. Lines transition to dashed below our convergence radius of $2.8\epsilon=320~\mathrm{pc}$. $V_{\mathrm{circ}}$ measurements at the half-light radius for observed MW satellites, as compiled in \cite{Silverman220310104}, are shown for the nine classical satellites ($L>10^5~L_{\mathrm{\odot}}$; solid error bars), and a subset of known ultrafaint dwarfs ($L<10^5~L_{\mathrm{\odot}}$; dashed error bars).}
    \label{fig:dwarf_density}
\end{figure*}

Figure~\ref{fig:dwarf_density} shows subhalo density profiles (left) and circular velocity profiles (right) in our CDM (top), SIDM (middle), and $T_{\mathrm{kd}}=1.46~\mathrm{keV}$ WSIDM (bottom) simulations. We restrict to the $30$ highest-$V_{\mathrm{peak}}$ subhalos because $V_{\mathrm{peak}}$ correlates with luminosity in our abundance-matching model; this cut therefore roughly selects the most easily detectable systems. We also impose $M_{\mathrm{sub}}>1.2\times 10^8~M_{\mathrm{\odot}}$ at $z=0$ before selecting these $30$ subhalos in each run. The color map indicates the abundance-matching-predicted luminosity, and the magenta line corresponds to the LMC analog in each simulation.

In CDM, we find a cuspy distribution of subhalo density profile slopes and a spread of $\mathcal{O}(1)~\mathrm{dex}$ about the mean inner density profile amplitude ($r\lesssim 1~\mathrm{kpc}$). Density profile amplitudes correlate with $V_{\mathrm{peak}}$, and thus with the predicted luminosity (which is set by $V_{\mathrm{peak}}$ in our abundance-matching model). On the other hand, in our SIDM simulation, high-mass subhalos have $\mathcal{O}(\mathrm{kpc})$ density cores, while subhalos with $M_{\mathrm{sub}}\lesssim 10^9~M_{\mathrm{\odot}}$ (i.e., $L \lesssim 10^5~L_{\mathrm{\odot}}$; \citealt{Nadler191203303}) have a range of density profile shapes, ranging from large cores to very cuspy inner profiles for core-collapsed objects. Thus, self-interactions diversify subhalo density profile amplitudes and slopes, thereby changing the relation between predicted densities and expected dwarf galaxy luminosities.

\begin{figure*}[t!]
\centering
\includegraphics[angle=270,origin=c,width=\textwidth,trim={-0.5cm 0cm 6cm 0cm}]{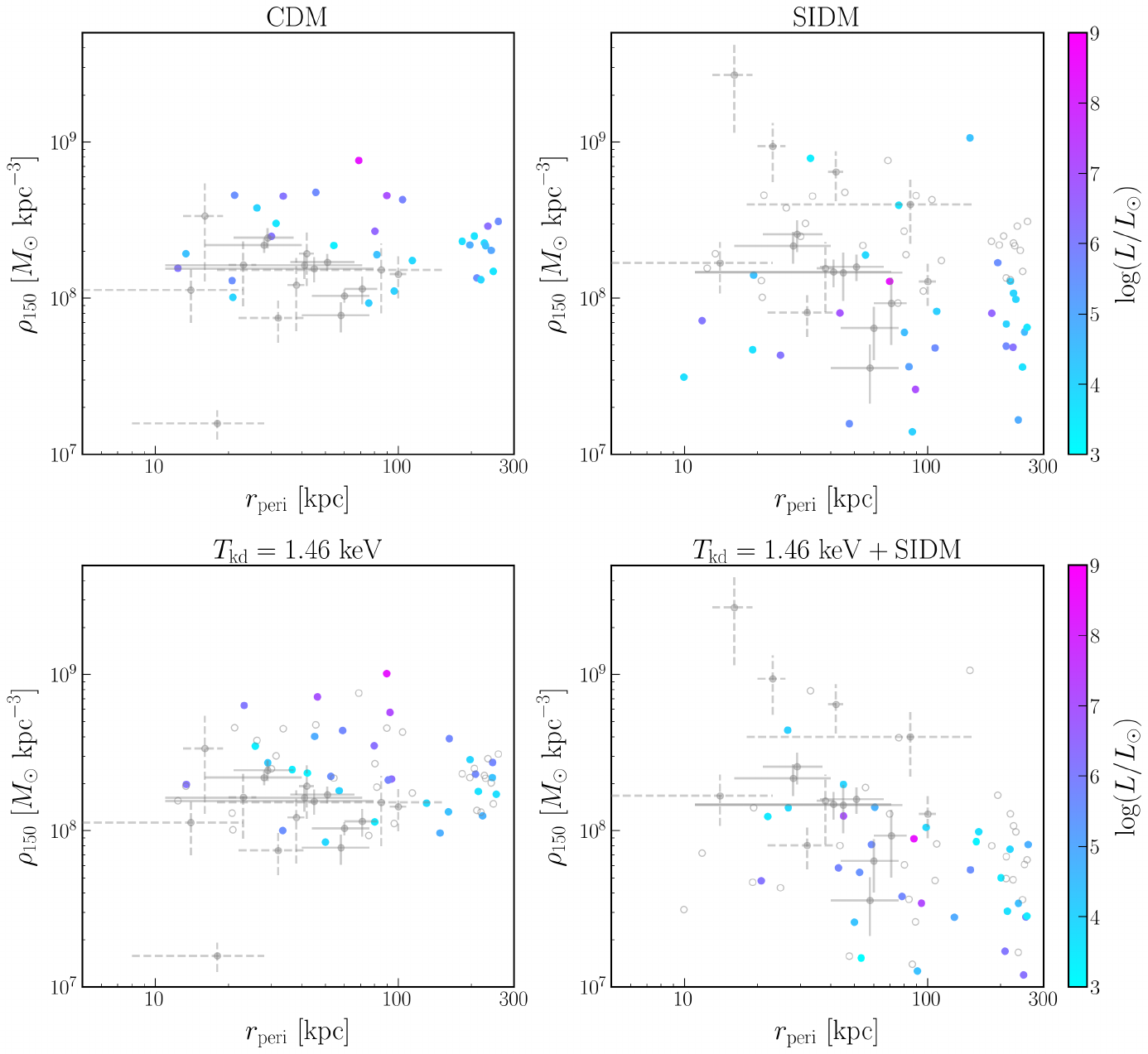}
    \caption{Distribution of central density evaluated at $150~\mathrm{pc}$, $\rho_{150}$, vs.\ pericentric distance, $r_{\mathrm{peri}}$, for the $30$ highest-$V_{\mathrm{peak}}$ subhalos with $M_{\mathrm{sub}}>1.2\times 10^8~M_{\mathrm{\odot}}$ in our CDM (top left), SIDM (top right), $T_{\mathrm{kd}}=1.46~\mathrm{keV}$--only (bottom left), and $T_{\mathrm{kd}}=1.46~\mathrm{keV}$ WSIDM (bottom right) simulations. Open gray circles in the top right and bottom left panels (bottom right panel) show the CDM (SIDM) result. Subhalos are colored by the luminosities of the dwarf galaxies they are predicted to host using the \cite{Nadler191203303} abundance-matching model. Derived $\rho_{150}$ and $r_{\mathrm{peri}}$ values for observed MW satellites, as compiled in \cite{Kaplinghat190404939}, are shown for the nine classical satellites ($L>10^5~L_{\mathrm{\odot}}$; solid error bars) and a subset of known ultrafaint dwarfs ($L<10^5~L_{\mathrm{\odot}}$; dashed error bars). These central densities are derived assuming Navarro--Frenk--White (NFW; \citealt{Navarro1997}) profiles appropriate for CDM and $T_{\mathrm{kd}}$--only models in the left panels, and isothermal profiles appropriate for SIDM and WSIDM models in the right panels.}
    \label{fig:rhoperi_rperi}
\end{figure*}

The bottom row of Figure~\ref{fig:dwarf_density} shows that introducing $P(k)$ suppression at the level probed by MW satellite abundances removes some of the lowest-mass subhalos that core collapse in our SIDM simulation, and erases the core-collapse signature in others. In addition, a few higher-mass subhalos that did not collapse in our SIDM simulation have very steep density profiles in WSIDM, indicating that changes to the merger and/or orbital histories caused by altering the ICs can accelerate core collapse for specific subhalos, even though it is inhibited in general for WSIDM systems (see Section~\ref{sec:f_cc}). Overall, the effect of $P(k)$ suppression on subhalo profiles is mild for the $T_{\mathrm{kd}}=1.46~\mathrm{keV}$ model. We show the $T_{\mathrm{kd}}=0.72~\mathrm{keV}$ WSIDM result in Appendix~\ref{sec:dwarf_dens_alt}, which is more extreme and results in only a few core-collapsed (sub)halos. Meanwhile, the distribution of subhalo profiles in our $T_{\mathrm{kd}}=2.32~\mathrm{keV}$ WSIDM simulation is very similar to that in SIDM.

These CDM, SIDM, and WSIDM results are reflected in the corresponding circular velocity profiles and are consistent with the CDM and SIDM results from \cite{Yang221113768}. For reference, we show circular velocity measurements for observed classical and ultrafaint MW satellites as compiled by \cite{Silverman220310104} in the circular velocity panels. The data are generally consistent with the high-amplitude cuspy profiles in our CDM simulation and the core-collapsed subhalos in our SIDM and WSIDM simulations. However, these comparisons are qualitative, and detailed forward modeling of satellites' observed stellar velocity dispersions and sizes, in the presence of a realistic MW disk potential, are needed to quantitatively constrain the SIDM and WSIDM models.

\subsection{Central Density-Pericenter Relation}
\label{sec:dwarf_peri}

Tidal stripping can accelerate gravothermal evolution (e.g., \citealt{Nishikawa190100499,Sameie190407872,Zeng211000259}). Thus, it is important to consider the orbital properties of observed satellites when interpreting their central densities in the context of SIDM and WSIDM. For example, \cite{Kaplinghat190404939} and \cite{Andrade231101528} reported an anticorrelation between observed satellites' orbital pericenters and central densities (however, see \citealt{Cardona-Barrero230406611}). To evaluate this relation for subhalos in our simulations, we measure $\rho_{150}$---the central density at a radius of $150~\mathrm{pc}$ from each subhalo's center---because many of the observed satellites' central densities have been measured at this radius (e.g., \citealt{Pace220505699}). Although $150~\mathrm{pc}$ is slightly below our simulations' formal convergence radius, we have tested that the correlations of interest are robust when the central densities are evaluated at $2.8\epsilon=320~\mathrm{pc}$ instead. We measure each subhalo's pericenter, $r_{\mathrm{peri}}$, as the smallest distance it achieves relative to the host across all snapshots using a cubic spline fit to subhalo and host three-dimensional positions. We again restrict to the $30$ highest-$V_{\mathrm{peak}}$ subhalos with $M_{\mathrm{vir}}>1.2\times 10^8~M_{\mathrm{\odot}}$ in each simulation.

Figure~\ref{fig:rhoperi_rperi} shows the $\rho_{150}$--$r_{\mathrm{peri}}$ relation in our CDM, SIDM, and $T_{\mathrm{kd}}=1.46~\mathrm{keV}$ simulations with and without self-interactions; we again restrict to the $30$ highest-$V_{\mathrm{peak}}$ subhalos in each case. These quantities are not significantly correlated in most of our runs. To quantify this, we calculate the Spearman rank correlation coefficient, $\rho$, which measures the correlation strength between ranked variables, with $+1$ ($-1$) representing perfect correlation (anticorrelation), and corresponding $p$-values. These tests yield $p>0.05$ in all cases except our $T_{\mathrm{kd}}=0.72$ and $1.46~\mathrm{keV}$ WSIDM runs. In these simulations, combining $P(k)$ suppression and self-interactions boosts the anticorrelation strength relative to SIDM or $T_{\mathrm{kd}}$--only runs, yielding $\rho=-0.47$ and $p=0.01$ in both cases. This boost is due to the presence of both core-forming and collapsing subhalos at low $r_{\mathrm{peri}}$, as well as a population of low-$\rho_{150}$ subhalos with large $r_{\mathrm{peri}}$. Similar results hold when comparing our CDM and SIDM simulations and our other $T_{\mathrm{kd}}$--only and WSIDM models, but our SIDM and $T_{\mathrm{kd}}=2.32~\mathrm{keV}$ WSIDM simulations do not produce statistically significant anticorrelations. See Appendix~\ref{sec:dwarf_peri_alt} for the $T_{\mathrm{kd}}=0.72$ and $2.32~\mathrm{keV}$ results.

Figure~\ref{fig:rhoperi_rperi} also shows observed classical and ultrafaint satellites' derived $\rho_{150}$ and $r_{\mathrm{peri}}$ values compiled in \cite{Kaplinghat190404939}. Satellites with both high and low central densities exist, particularly at small $r_{\mathrm{peri}}$; this is even more noticeable when their density profiles are inferred using an isothermal profile. SIDM and WSIDM qualitatively reproduce the observed central density diversity due to the existence of both cored and core-collapsed subhalos. In contrast, \cite{Ebisu210705967} show that velocity-independent SIDM with $\sigma_0/m_{\chi}\sim 1$ to $3~\mathrm{cm^2~g}^{-1}$ uniformly reduces the subhalos' central densities, rather than mimicking the observed spread in the $\rho_{150}$--$r_{\mathrm{peri}}$ relation, because core collapse does not occur in these scenarios. Similar to Figure~\ref{fig:dwarf_density}, quantitative comparisons to our simulation predictions will require forward modeling the satellites' velocity dispersions and orbits, including measurement systematics. In parallel, it will be important to scrutinize our results using improved subhalo finders that track highly stripped subhalos (e.g., \textsc{Symfind}; \citealt{Mansfield230810926}) and by running SIDM and WSIDM extensions of embedded disk simulation suites (e.g., EDEN; \citealt{Wang240801487}). In particular, the MW disk may disrupt core-forming subhalos with small pericentric distances, thereby increasing the anticorrelation central density--pericenter anticorrelation; we return to this point in Section~\ref{sec:caveats}.

\section{Discussion}
\label{sec:discussion}

We now discuss our results, focusing on the implications of $P(k)$ suppression for core collapse in our WSIDM simulations (Section~\ref{sec:cc_realistic}), comparison to previous work and particularly ETHOS simulations (Section~\ref{sec:comparison}), and areas for future work associated with our setup and analyses (Section~\ref{sec:caveats}).

\subsection{Core Collapse in Warm Self-interacting Dark Matter}
\label{sec:cc_realistic}

Core collapse is a distinct signature of SIDM models. If DM self-interactions significantly impact the inner density profiles of subhalos that host MW satellite galaxies, core collapse is needed in a significant fraction of these systems to avoid underpredicting their central densities (e.g., \citealt{Silverman220310104}). As discussed in Section~\ref{sec:dwarf_peri}, this conclusion is strengthened when considering the orbital properties of observed MW satellites. In particular, SIDM models that uniformly core MW subhalos, rather than producing core collapse in a fraction of these systems, yield subhalo populations that are not sufficiently diverse to account for the observed distribution of MW satellite central densities and pericenters (e.g., \citealt{Ebisu210705967}). This situation motivates a strong, velocity-dependent cross section \citep{Correa200702958,Yang221113768}. Similar conclusions hold when the dwarfs' stellar velocity dispersions are modeled directly (e.g., \citealt{Kim210609050}).

For a wide range of SIDM models encapsulated by Equation~\ref{eq:lagrangian}, the velocity-dependent cross sections with large $\sigma_0/m_{\chi}$ needed to achieve core collapse are naturally accompanied by a cutoff in $P(k)$. This cutoff affects the formation and evolution of low-mass (sub)halos, which are precisely the systems expected to core collapse in the absence of $P(k)$ suppression. By simulating three benchmark models with varying levels of $P(k)$ suppression, we find that WSIDM (sub)halo population statistics vary depending on the amplitude of $P(k)$ suppression; furthermore, WSIDM can differentially impact isolated halos versus subhalos relative to SIDM. Focusing on the core-collapsed population, we highlight the following conclusions from our results.
\begin{itemize}
    \item In the model featuring extreme $P(k)$ suppression ($T_{\mathrm{kd}}=0.72~\mathrm{keV}$), roughly half of the (sub)halos that collapse in our SIDM simulation never form (Figures~\ref{fig:shmf} and \ref{fig:tau_0_vmax_alt}); the remaining systems' growth histories are significantly delayed and suppressed. These effects combine to nearly entirely remove the core-collapse signature.
    \item As the $P(k)$ cutoff moves to smaller scales ($T_{\mathrm{kd}}=1.46~\mathrm{keV}$), (S)HMF suppression weakens (Figure~\ref{fig:shmf}). However, the core-collapsed fraction is still significantly reduced because low-mass (sub)halo growth is suppressed (and mildly delayed) relative to simulations without a $P(k)$ cutoff (Figures~\ref{fig:tau_0} and \ref{fig:matched_halos}).
    \item For a mild $P(k)$ cutoff ($T_{\mathrm{kd}}=2.32~\mathrm{keV}$), the (S)HMF is weakly suppressed (Figure~\ref{fig:shmf}), but the core-collapsed population is moderately suppressed relative to SIDM (Figure~\ref{fig:tau_0}). This illustrates the extreme sensitivity of gravothermal evolution to (sub)halo growth histories.
\end{itemize} 

In all cases, this reduction of the population of core-collapsed systems is more severe for isolated halos compared to subhalos (e.g., see Figure~\ref{fig:tau_0}), likely because the subhalos' gravothermal evolution is accelerated by tidal stripping whereas the isolated halos' gravothermal evolution is (to first order) determined by their accretion of lower-mass objects, which probe $P(k)$ on smaller scales.

\subsection{Comparison to Previous Work}
\label{sec:comparison}

Several authors have considered the direct impact of DM self-interactions on $P(k)$. For example, \cite{Atrio-Barandela9702236} and \cite{Hannestad0003046} found that certain SIDM models (e.g., involving a massive vector boson mediator) slightly reduce small-scale $P(k)$ suppression relative to a corresponding WDM model (also see \citealt{Atrio-Barandela0106228}). Recent works that consider a wider range of SIDM models reach similar conclusions in linear theory (e.g., \citealt{Huo190702454,Yunis200205778,Yunis210802657,Egana-Ugrinovic210206215,Garani220106551}). In our WSIDM models, power spectrum suppression is slightly steeper than in the corresponding thermal-relic WDM models with equal half-mode scales, and there are low-amplitude DAOs below the initial cutoff. Both features are consistent with the general picture outlined by previous works.

To our knowledge, the only previous simulations that include both $P(k)$ suppression and DM self-interactions were also run in the ETHOS framework \citep{Cyr-Racine151205344}. We compare to \cite{Vogelsberger151205349}, since their simulation resolution and zoom-in setup is broadly similar to ours. These authors consider ETHOS models that roughly match the half-mode scales of $1.89$, $2.67$, and $3.66~\mathrm{keV}$ thermal-relic WDM, with SIDM cross sections that are comparable to our MilkyWaySIDM model. Thus, their least suppressed model is similar to our most suppressed $T_{\mathrm{kd}}=0.72~\mathrm{keV}$ scenario, which is half-mode matched to $m_{\mathrm{WDM}}=3.5~\mathrm{keV}$. As we have shown, $P(k)$ suppression is so strong in this model that core collapse is almost entirely eliminated, even though it would be achieved with CDM ICs. This is due to both the erasure of low-mass subhalos and their delayed and suppressed growth. Subhalo maximum circular velocity functions in \cite{Vogelsberger151205349} are uniformly suppressed relative to a CDM simulation without self-interactions, consistent with this interpretation. Our study therefore complements \cite{Vogelsberger151205349} by including models with sufficiently mild $P(k)$ suppression to preserve core collapse.

\subsection{Future Work}
\label{sec:caveats}

We have presented DM--only zoom-in simulations of a MW analog in CDM, SIDM, and WSIDM. There are several areas for future work and potential caveats: host-to-host variance, subhalo finding, convergence of the (sub)halos' gravothermal evolution, and the impact of baryons.

First, regarding host-to-host variance, we expect WSIDM (S)HMF suppression relative to CDM to be similar across MW zoom-in environments because it is determined by the small-scale $P(k)$ cutoff. For example, \citetalias{Nadler241003635} showed that WDM SHMF suppression is consistent across three different MW hosts. Predictions for absolute (sub)halo abundances depend on host mass, accretion history, and environment; modeling WSIDM effects in large samples of MW analogs will therefore be important. We note that \cite{Meshveliani231200150} measured an environmental dependence of WDM halo mass function suppression across a cosmological volume that includes underdense regions, which our zoom-ins avoid by construction. It will be interesting to model this effect in the Local Volume using constrained simulations \citep{Carlesi160203919,McAlpine220204099,Wempe240602228}.

Second, regarding subhalo finding, we have used the \textsc{Rockstar} halo finder and \textsc{consistent-trees} merger tree codes \citep{Behroozi11104372,Behroozi11104370}, which are known to prematurely lose track of certain subhalos in CDM. This effect is most severe for highly stripped subhalos near hosts' centers, but can persist even for (initially) massive systems \citep{Mansfield230810926}. In SIDM and WSIDM, many subhalos are cored at infall, and it is possible for tides to completely unbind these systems (e.g., \citealt{Errani221001131}). On the other hand, it is more challenging to reliably identifying cored versus cuspy subhalos, although this has not been thoroughly studied. Thus, it is possible that the $\mathcal{O}(10\%)$ subhalo abundance suppression we measure due to self-interactions (Figure~\ref{fig:shmf}, right panel) is partly due to incomplete subhalo finding. Nonetheless, we demonstrate that our main results are converged in Appendix~\ref{sec:convergence}, and we have focused on comparisons between SIDM and WSIDM that mitigate this systematic.

Third, regarding convergence of our SIDM and WSIDM (sub)halo density profiles, recent studies highlight a set of related challenges: finite particle resolution can add artificial scatter to predicted collapse times, extremely low-concentration or core-collapsed subhalos can be systematically mis-modeled, and the self-interaction algorithm and/or simulation parameters can bias predictions \citep{Zhong230608028,Fischer240300739,Mace240201604,Palubski240212452}. We find very few of the extremely low-concentration (sub)halos that \cite{Mace240201604} show are particularly difficult to simulate; meanwhile, our core-collapsed (sub)halos' central densities are typically enhanced relative to CDM by an $\mathcal{O}(1)$ factor, which implies that these systems are not in the deeply core-collapsed phase where simulations can fail to conserve energy \citep{Fischer240300739,Palubski240212452}. However, most (sub)halos in our SIDM and WSIDM have a dimensionless scattering cross section $\hat{\sigma} \equiv (\sigma/m_{\chi})\rho_s r_s \lesssim 0.1$ (where $\rho_s$ and $r_s$ are, respectively, the scale density and radius of the initial NFW profile; \citealt{Mace240201604}), and are therefore in the long mean free path regime. Furthermore, our SIDM scattering algorithm has been calibrated against high-resolution controlled simulations \citep{Yang220503392} and against halos with ICs drawn from the cosmological parent box used to initialize our zoom-in simulations (see \citealt{Yang221113768}, Appendix B). Furthermore, we use a time-stepping parameter of $\eta=0.01$, which is conservative for cosmological simulations. Thus, we expect discreteness noise to be the dominant systematic for our gravothermal evolution predictions, since we analyze (sub)halo density profiles down to $2000$ particles, whereas \cite{Palubski240212452} and \cite{Mace240201604} show that even a resolution of $\sim 10^5$ particles leads to artificial scatter. It will therefore be important to assess the scatter in our (sub)halo density profiles predictions using higher-resolution simulations and semianalytic models.

Finally, we discuss the impact of baryons. In the presence of baryons, the cored host density profiles in our SIDM and WSIDM simulations (Appendix~\ref{sec:host_density}) will transform into cusps (e.g., \citealt{Kaplinghat13116524,Sameie180109682,Robles190301469,Rose220614830,Correa240309186}), increasing the severity of tidal stripping and disruption. Thus, our SIDM and WSIDM SHMF suppression predictions in Figure~\ref{fig:shmf} are likely underestimates. The presence of a central galaxy also enhances tidal stripping within a given DM model \citep{Wang240801487}, which may further diversify the profiles of SIDM and WSIDM subhalos that orbit near the host center. In particular, low-density SIDM and WSIDM subhalos (see Figures~\ref{fig:dwarf_density} and \ref{fig:rhoperi_rperi}) may be disrupted by the disk, potentially bringing our predictions into better agreement with the data. Finally, baryons within (sub)halos may alter their density profiles via supernova feedback (e.g., \citealt{Pontzen11060499,Read180806634}). This effect is likely negligible for ultrafaint dwarfs, but it affects classical and bright dwarfs \citep{Tollet150703590,Lazar200410817}. Furthermore, in SIDM, the DM distribution responds efficiently to baryons, which changes the core-collapse timescale (e.g., \citealt{Feng201015132,Zhong230608028,Yang240503787}). Thus, it is timely to develop hydrodynamic SIDM and WSIDM simulations that combine strong, velocity-dependent self-interactions and stellar feedback (e.g., following \citealt{Despali250112439}).


\section{Summary and Outlook}
\label{sec:conclusions}

We have presented eight cosmological zoom-in simulations of a MW analog as the third installment of the COZMIC suite, focusing on the combined effects of $P(k)$ suppression and SIDM (Figure~\ref{fig:transfer}). For the first time, our simulations capture the full range of (sub)halos' gravothermal evolution, including core collapse. We found that $P(k)$ suppression and self-interactions predicted in favored SIDM models can simultaneously affect (sub)halo abundances and density profiles. Jointly modeling these effects will therefore be critical for probing WSIDM models with small-scale structure data.

Our main results are as follows.
\begin{enumerate}
    \item (S)HMF suppression is determined by the suppression in the linear matter power spectrum $P(k)$ used to generate the ICs, rather than late-time self-interactions. The suppression is set by the kinetic decoupling temperature in our WSIDM model (Figure~\ref{fig:shmf});
    \item Power spectrum suppression reduces the SIDM core-collapse signature. This reduction is more severe when $P(k)$ is more suppressed, and is more pronounced for isolated halos versus subhalos (Figures~\ref{fig:tau_0_vmax} and \ref{fig:tau_0});
    \item This reduction of the core-collapsed population is mainly due to the (sub)halos' suppressed growth in WSIDM, which lowers their initial concentrations and increases their core-collapse timescales (Figure~\ref{fig:matched_halos}); 
    \item Subhalo density profiles are diversified by the SIDM model we consider, but are not significantly affected when additionally introducing $P(k)$ suppression consistent with current bounds (Figure~\ref{fig:dwarf_density}).
    \item WSIDM models with mild $P(k)$ suppression can produce an anticorrelation between subhalo central density and pericentric distance due to core collapse in subhalos with small pericenters and core formation in subhalos with large pericenters (Figure~\ref{fig:rhoperi_rperi}); these models also increase the abundance of low-concentration isolated systems relative to SIDM--only predictions.
\end{enumerate}

This work highlights the importance of combining dwarf galaxy luminosity functions and density profile measurements to simultaneously constrain early and late-time DM physics. Furthermore, our WSIDM results reveal a possible new-physics discovery scenario that is not available within SIDM models without $P(k)$ suppression. In WSIDM, $P(k)$ suppression is proportional to the SIDM coupling strength. Thus, models with stronger self-interactions produce more severe $P(k)$ suppression, which partially or even entirely removes the core-collapse signature, as we show in this study. This degeneracy can be broken using observations that probe the (S)HMF, which provides an independent constraint on $P(k)$ suppression. In other words, there is a finite range of WSIDM parameter space where a sizable fraction of dwarf galaxies both form and core collapse. Future efforts to map out this parameter space can complement previous studies that combine $P(k)$ suppression and SIDM core formation \citep{Vogelsberger151205349,Drlica-Wagner190201055}.

In the context of dwarf galaxies, these results imply that both photometric surveys, which can probe dwarf galaxy abundances throughout the Local Volume, and spectroscopic facilities, which can probe the halo density profiles of faint systems, are critical for testing DM self-interaction physics (see \citealt{Chakrabarti220306200} for an overview of future observational facilities for studying DM). This synergy will be important as the Rubin Observatory Legacy Survey of Space and Time (\citealt{Ivezic08052366}) begins to discover large numbers of faint dwarf galaxies in the coming years (e.g., \citealt{Mutlu-Pakdil210501658,Tsiane240416203}), increasing the precision of $P(k)$ constraints \citep{Nadler240110318}. Combining these data with other observational probes of low-mass (sub)halos, including strong gravitational lenses and stellar streams, will enable powerful tests of DM self-interactions.


\section*{Acknowledgements}

Halo catalogs, merger trees, and particle snapshots are distributed in Zenodo at doi: 10.5281/zenodo.14666735. Analysis code is available at \url{https://github.com/eonadler/COZMIC/}.

We thank Francis-Yan Cyr-Racine for helpful discussions. This work was supported by the John Templeton Foundation under grant ID \#61884 and the U.S. Department of Energy (DoE) under grant No.\ de-sc0008541 (D.Y.\ and H.-B.Y.). The opinions expressed in this publication are those of the authors and do not necessarily reflect the views of the John Templeton Foundation. V.G.\ acknowledges the support from NASA through the Astrophysics Theory Program, Award Number 21-ATP21-0135, the National Science Foundation (NSF) CAREER grant No. PHY2239205, and from the Research Corporation for Science Advancement under the Cottrell Scholar Program. This research was supported in part by grant NSF PHY-2309135 to the Kavli Institute for Theoretical Physics (KITP).

The computations presented here were conducted through Carnegie's partnership in the Resnick High Performance Computing Center, a facility supported by Resnick Sustainability Institute at Caltech. Computations were also performed using the clusters and data storage resources of the HPCC at UCR, which were funded
by grants from NSF (MRI-2215705 and MRI-1429826) and
NIH (1S10OD016290-01A1). This
work used data from the Milky Way-est suite of simulations, hosted at \url{https://web.stanford.edu/group/gfc/gfcsims/}, which was supported by the Kavli Institute for Particle Astrophysics and Cosmology at Stanford, SLAC National Accelerator Laboratory, and the U.S.\ DoE under contract number DE-AC02-76SF00515 to SLAC.

\software{
{\sc consistent-trees} \citep{Behroozi11104370},
\textsc{Helpers}, (\http{bitbucket.org/yymao/helpers/src/master/}),
\textsc{Jupyter} (\http{jupyter.org}),
\textsc{Matplotlib} \citep{matplotlib},
\textsc{NumPy} \citep{numpy},
\textsc{pynbody} \citep{pynbody},
{\sc Rockstar} \citep{Behroozi11104372},
\textsc{SciPy} \citep{scipy},
\textsc{Seaborn} (\https{seaborn.pydata.org}).
}

\bibliographystyle{yahapj2}
\bibliography{references,software}


\appendix

\begin{figure*}[t!]
\centering
\includegraphics[width=0.49\textwidth]{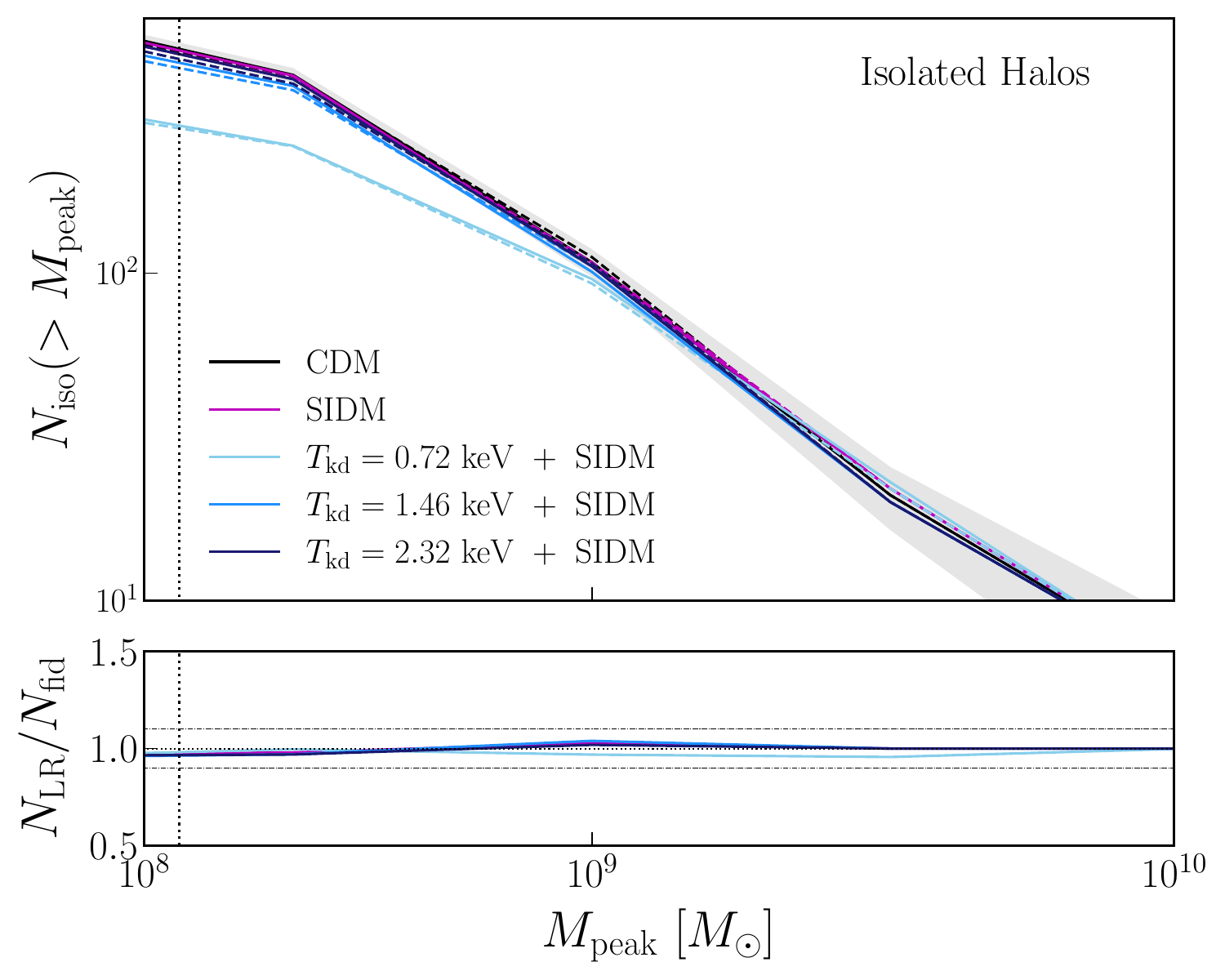}
\includegraphics[width=0.49\textwidth]{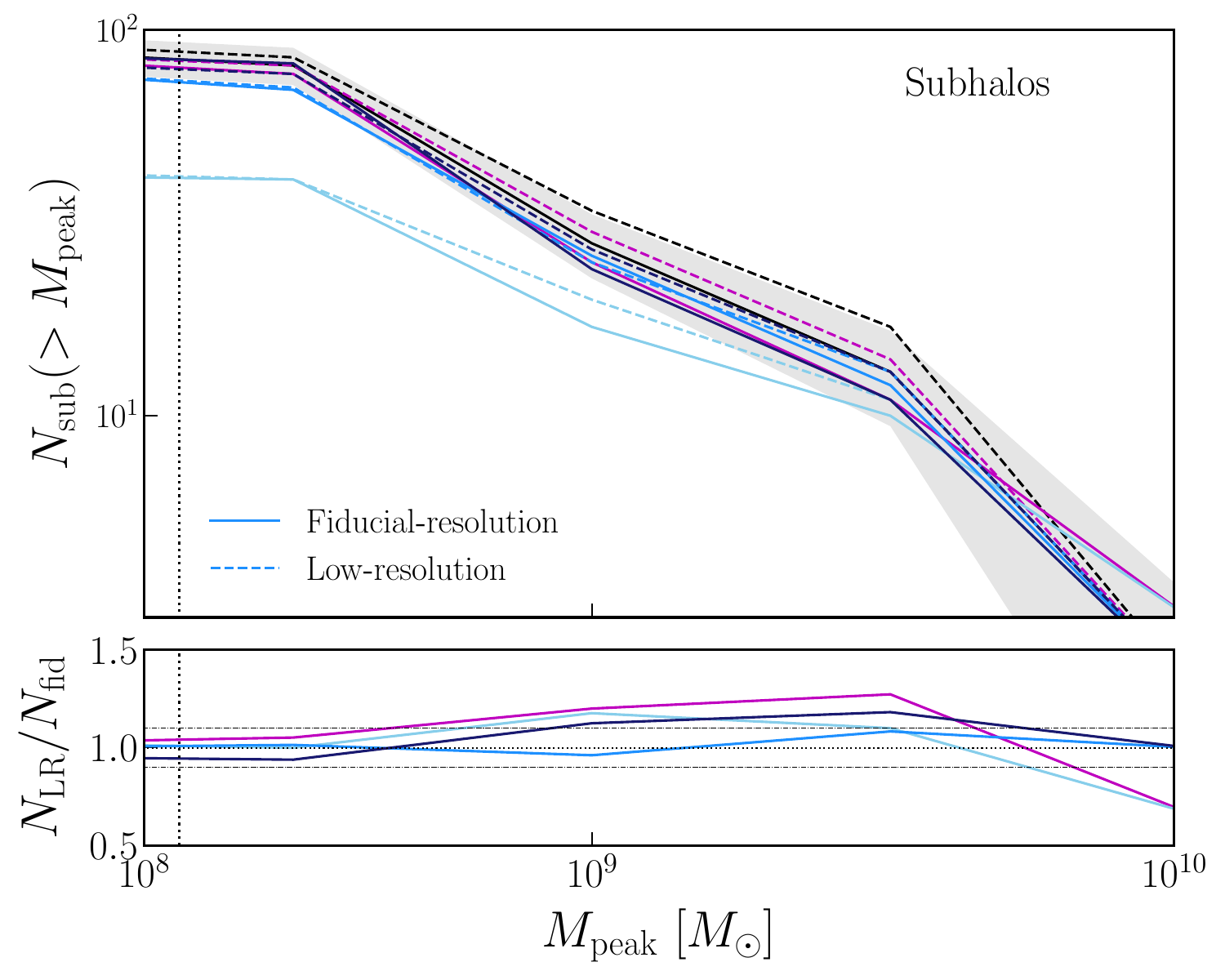}
    \caption{Same as Figure~\ref{fig:shmf}, but comparing our fiducial-resolution (solid) and LR (dashed) simulations to assess convergence. Both resolution levels are subject to a $M_{\mathrm{vir}}>1.2\times 10^8~M_{\mathrm{\odot}}$ cut, corresponding to $300$ particles in the LR simulations and $2400$ particles in the fiducial-resolution simulations. The vertical dotted line shows the $300$ particle limit for the LR simulations. The horizontal dotted--dashed lines in the bottom panels show $\pm 10\%$ deviations from the fiducial-resolution result.}
    \label{fig:shmf_convergence}
\end{figure*}

\section{Convergence Tests}
\label{sec:convergence}

To assess convergence, we resimulate all models using four \textsc{MUSIC} refinement regions, yielding a high-resolution DM particle mass of $4\times 10^5~M_{\mathrm{\odot}}$; we use a softening of $\epsilon=170~\mathrm{pc}~h^{-1}$ for these ``LR'' runs. Note that our LR runs have equivalent resolution the the ``fiducial'' simulations in \citetalias{Nadler241003635}, while our fiducial-resolution simulations have equivalent resolution to the ``high-resolution'' runs in \citetalias{Nadler241003635}. Below, we compare our LR and fiducial-resolution simulations to analyze (S)HMF (Appendix~\ref{sec:hmf_convergence}) and $R_{\mathrm{max}}$--$V_{\mathrm{max}}$ relation (Appendix~\ref{sec:rmax_vmax_convergence}) convergence.

\subsection{Mass Functions}
\label{sec:hmf_convergence}

In \citetalias{Nadler241003635}, we showed that SHMF suppression is converged across the zoom-in hosts and beyond-CDM models for simulations with identical settings to ours. Here, we repeat this analysis for a single host, because we only simulate one MW--like system in this work, and we focus on our SIDM and WSIDM results because convergence of SHMF suppression due to $P(k)$ suppression alone was already demonstrated in \citetalias{Nadler241003635}.

Figure~\ref{fig:shmf_convergence} shows the isolated-halo (left) and subhalo (right) peak-mass function in our LR and fiducial-resolution simulations. We impose a present-day virial mass cut corresponding to $300$ LR particles and $2400$ fiducial-resolution particles, i.e., $M_{\mathrm{vir}}>1.2\times 10^8~M_{\mathrm{\odot}}$. Isolated-halo mass functions are converged at the percent level above this threshold. Subhalo abundances are less well converged, but still only deviate from the fiducial-resolution result by at most $\approx 10\%$; at high subhalo masses, Poisson uncertainties are large and we sometimes observe enhancements in LR subhalo abundances. Thus, our fiducial-resolution simulations resolve the isolated halo and SHMF well for $M_{\mathrm{vir}}>1.2\times 10^8~M_{\mathrm{\odot}}$, which is the cut we impose in our main analyses related to subhalo profiles. Convergence down to the $300$ particle limit in our fiducial-resolution simulations, i.e., $M_{\mathrm{vir}}>1.5\times 10^7~M_{\mathrm{\odot}}$, will need to be tested using even higher-resolution simulations in future work. However, none of our main conclusions rely on (sub)halos in this regime.

\subsection{$R_{\mathrm{max}}$--$V_{\mathrm{max}}$ Relations}
\label{sec:rmax_vmax_convergence}

\begin{figure*}[t!]
\centering
\includegraphics[width=\textwidth]{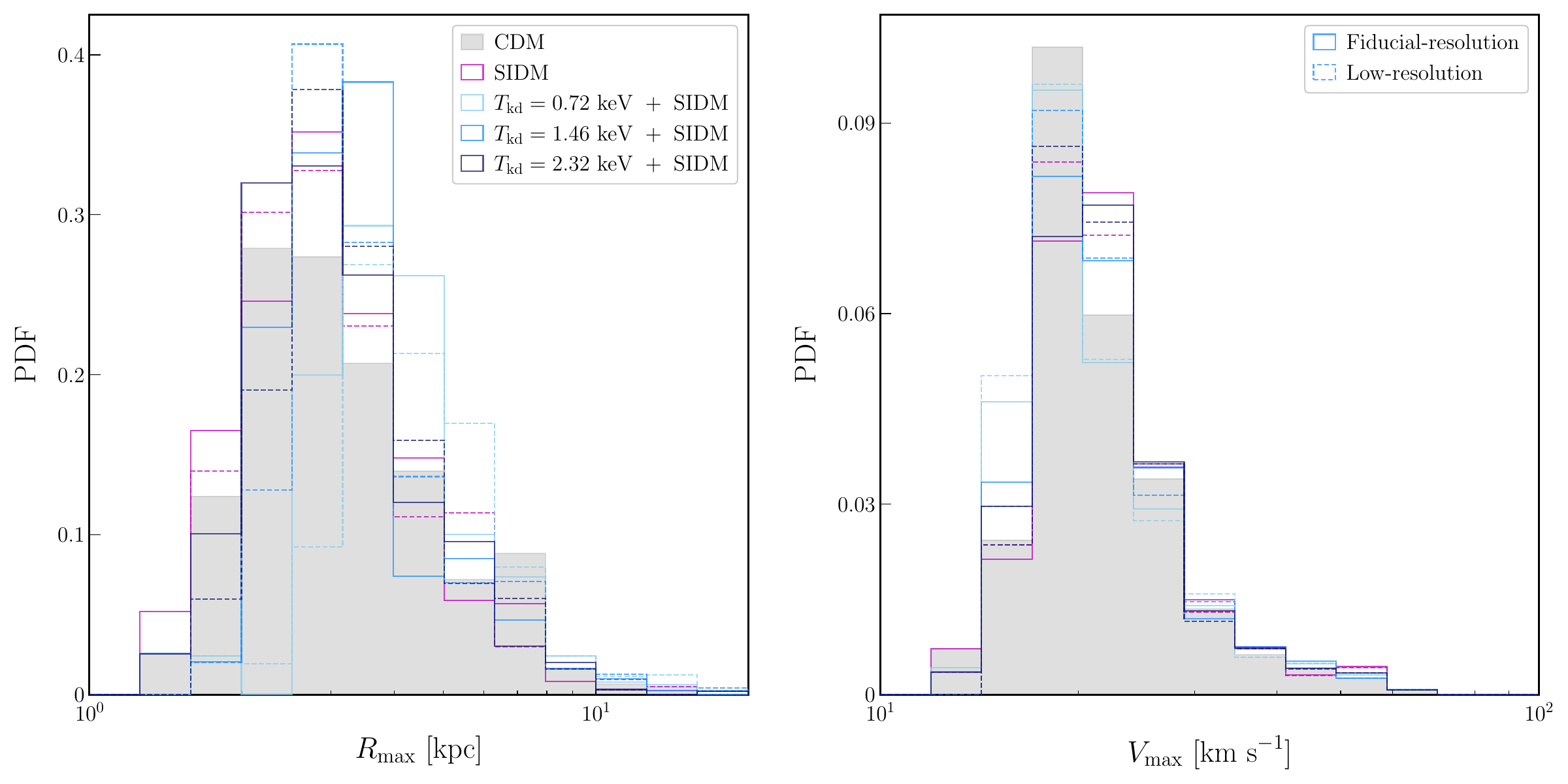}
    \caption{Normalized distributions of $R_{\mathrm{max}}$ (left) and $V_{\mathrm{max}}$ (right) for isolated halos with $M_{\mathrm{vir}}>8\times 10^8~M_{\mathrm{\odot}}$ in our CDM (black), SIDM (magenta), and $T_{\mathrm{kd}}=0.72$, $1.46$, and $2.32~\mathrm{keV}$ WSIDM simulations (light to dark blue). Solid (dashed) lines show the fiducial-resolution LR result.}
    \label{fig:rmax_vmax_convergence}
\end{figure*}

To assess convergence of $R_{\mathrm{max}}$ and $V_{\mathrm{max}}$, Figure~\ref{fig:rmax_vmax_convergence} shows $R_{\mathrm{max}}$ (left) and $V_{\mathrm{max}}$ (right) distributions for isolated halos. We impose a mass cut corresponding to $2000$ LR particles and 16,000 fiducial-resolution particles, i.e., $M_{\mathrm{vir}}>8\times 10^8~M_{\mathrm{\odot}}$; this yields too few subhalos to measure the distributions reliably, so we focus on isolated halos. All pairs of LR and fiducial-resolution distributions are statistically consistent: two-sample Kolmogorov–Smirnov (K-S) tests yield $p\gg 0.05$ in all cases. The $V_{\mathrm{max}}$ distributions are extremely similar, with $p>0.99$ from all two-sample K-S tests, while K-S tests yield $p\approx 0.3$ for all $R_{\mathrm{max}}$ distribution comparisons.

Thus, our fiducial-resolution simulations resolve isolated halo $R_{\mathrm{max}}$ and $V_{\mathrm{max}}$ distributions well for $M_{\mathrm{vir}}>8\times 10^8~M_{\mathrm{\odot}}$. A larger simulation suite is needed to statistically assess convergence for subhalos, and higher-resolution runs are needed to assess convergence in the $M_{\mathrm{vir}}<8\times 10^8~M_{\mathrm{\odot}}$ regime. Nonetheless, by inspecting the $R_{\mathrm{max}}$ and $V_{\mathrm{max}}$ evolution histories for individual pairs of (sub)halos matched between our LR and fiducial-resolution runs, we do not find evidence that our results are biased in this regime.

\section{Comparison to Warm Dark Matter (Sub)halo Mass Function Suppression}
\label{sec:wdm_comparison}

\begin{figure*}[t!]
\centering
\includegraphics[width=0.49\textwidth]{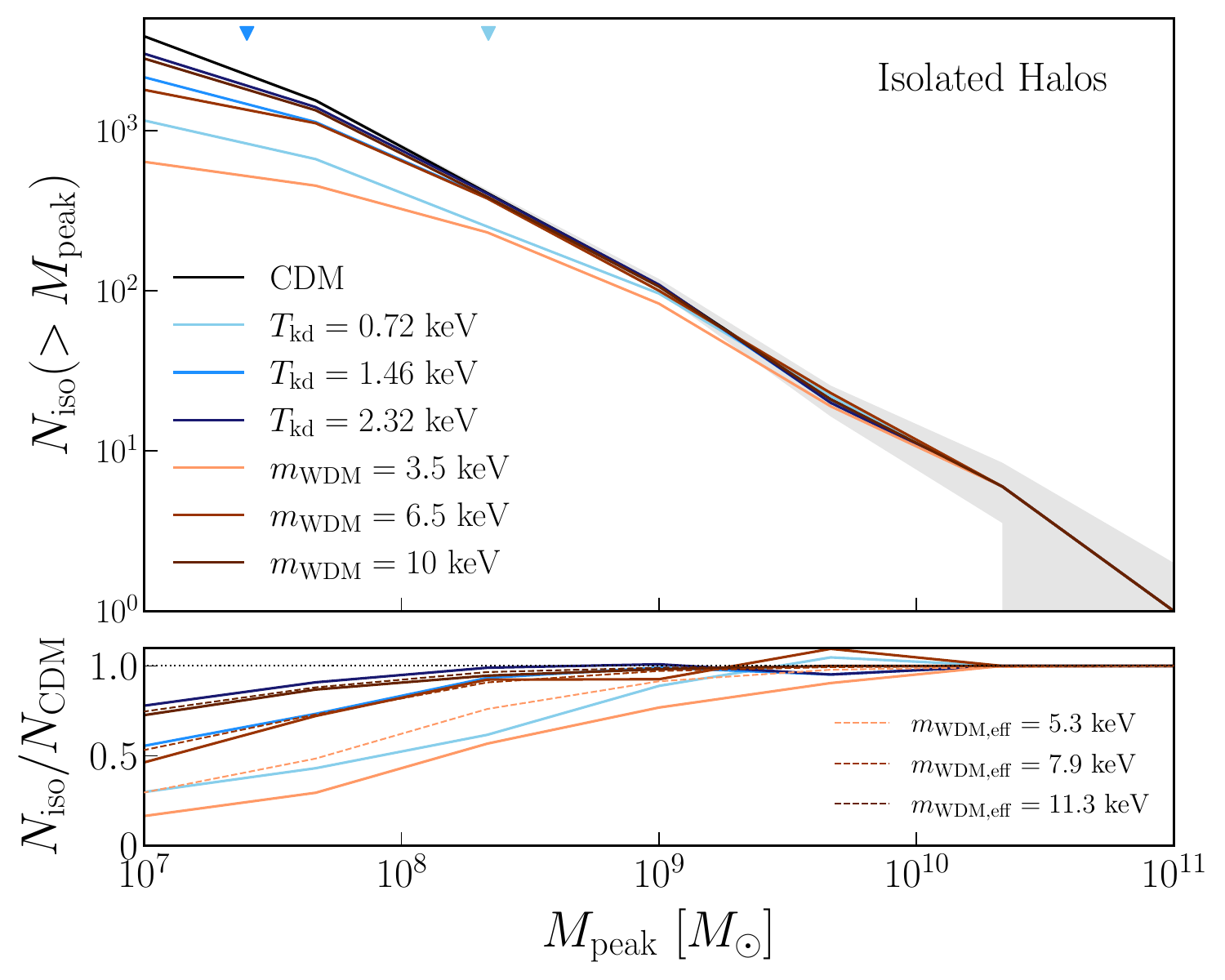}
\includegraphics[width=0.49\textwidth]{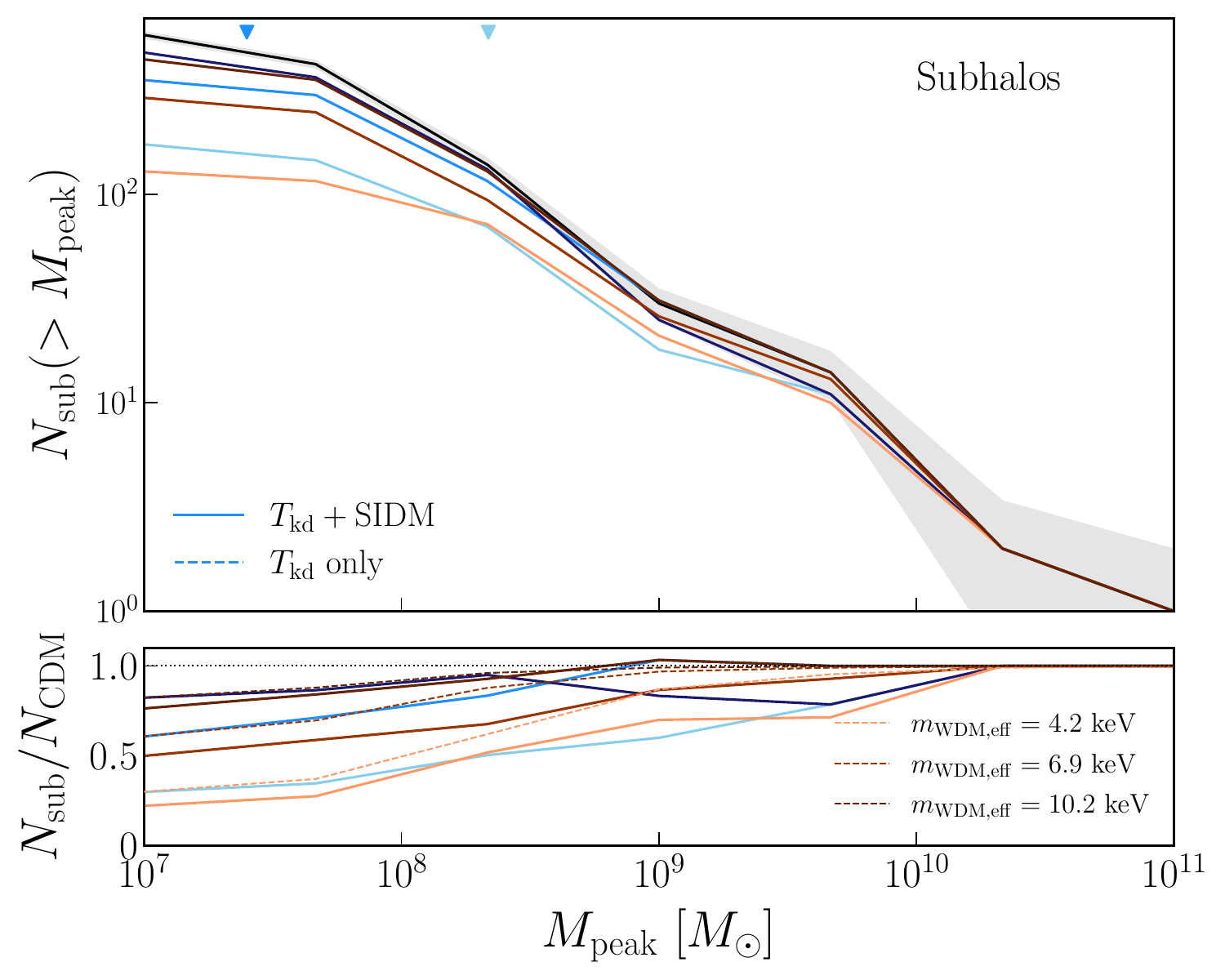}
    \caption{Same as Figure~\ref{fig:shmf}, but comparing our $T_{\mathrm{kd}}=0.72$, $1.46$, and $2.32~\mathrm{keV}$--only isolated-halo (left) and subhalo (right) mass functions (light to dark blue) to our $m_{\mathrm{WDM}}=3.5$, $6.5$, and $10~\mathrm{keV}$ simulations (light to dark red). Dashed lines in the bottom panels show ``effective'' WDM models matched to each of our $T_{\mathrm{kd}}$--only runs.}
    \label{fig:shmf_wdm}
\end{figure*}

Figure~\ref{fig:shmf_wdm} compares isolated halo and SHMFs in our fiducial-resolution $T_{\mathrm{kd}}$--only and half-mode-matched WDM simulations, using our usual $M_{\mathrm{vir}}>1.5\times 10^7~M_{\mathrm{\odot}}$ cut in all cases. The $T_{\mathrm{kd}}$--only mass functions are less suppressed than the half-mode-matched WDM results. This is similar to our comparison between the DM--baryon scattering and WDM models in \citetalias{Nadler241003635}, where we showed that the reduction in SHMF suppression for interacting DM models is largely due to the difference in the shape of the initial $P(k)$ cutoff rather than DAOs. Thus, we expect the low-amplitude DAOs in our $T_{\mathrm{kd}}$ models to have a small impact on (sub)halo abundances given our $M_{\mathrm{vir}}$ cut, although we note that the difference between our $T_{\mathrm{kd}}$--only and half-mode-matched WDM mass functions is largest for our $T_{\mathrm{kd}}=0.72~\mathrm{keV}$--only run, which has a DAO that is fully resolved in our simulations when mapping wavenumbers to masses in linear theory (see the right panel of Figure~\ref{fig:transfer}).

The bottom panels of Figure~\ref{fig:shmf_wdm} show mass function ratios relative to CDM for our $T_{\mathrm{kd}}$--only and half-mode-matched WDM simulations; we also show results for the ``effective'' WDM models used to assess the impact of low-mass (sub)halo erasure on gravothermal evolution in Section~\ref{sec:f_cc}. We derive these ``effective'' WDM models following the procedure for DM--baryon scattering simulations in \citetalias{Nadler241003635}. In particular, we apply the best-fit WDM SHMF suppression model derived in \citetalias{Nadler241003635} to our CDM simulations, and we find the ``effective'' WDM mass, $m_{\mathrm{WDM,eff}}$, that matches the total number of isolated halos (or subhalos) with $M_{\mathrm{vir}}>1.5\times 10^7~M_{\mathrm{\odot}}$ in each of our $T_{\mathrm{kd}}$--only simulations. Thus, the ``effective'' WDM curves in the bottom panels of Figure~\ref{fig:shmf_wdm} closely match the $T_{\mathrm{kd}}$--only results at $M_{\mathrm{peak}}=1.5\times 10^7~M_{\mathrm{\odot}}$ by construction, and the differences in the shape of the suppression at higher (sub)halo masses reflect differences in the underlying shapes of $P(k)$.

\section{Matched Subhalo Evolution}
\label{sec:matched_alt}

\begin{figure*}[t!]
\hspace{-25mm}
\includegraphics[width=1.25\textwidth]{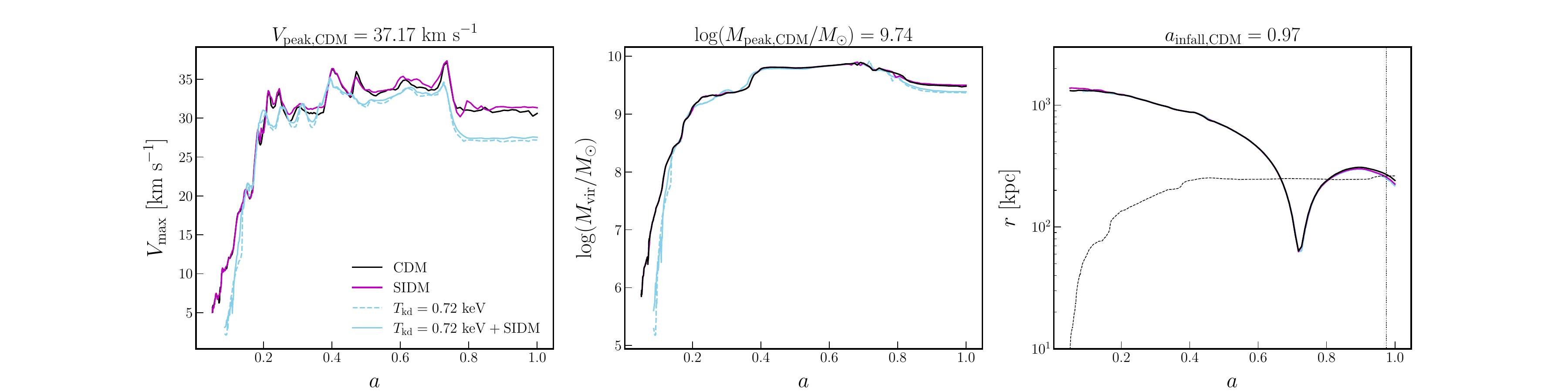} \\
\hspace*{-25mm}
\includegraphics[width=1.25\textwidth]{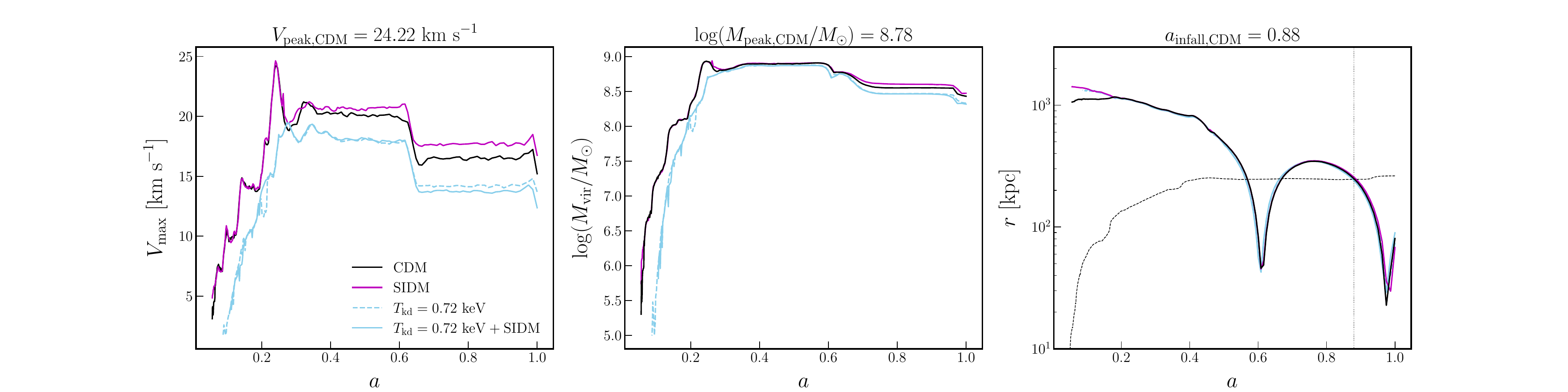} \\
\hspace*{-25mm}
\includegraphics[width=1.25\textwidth]{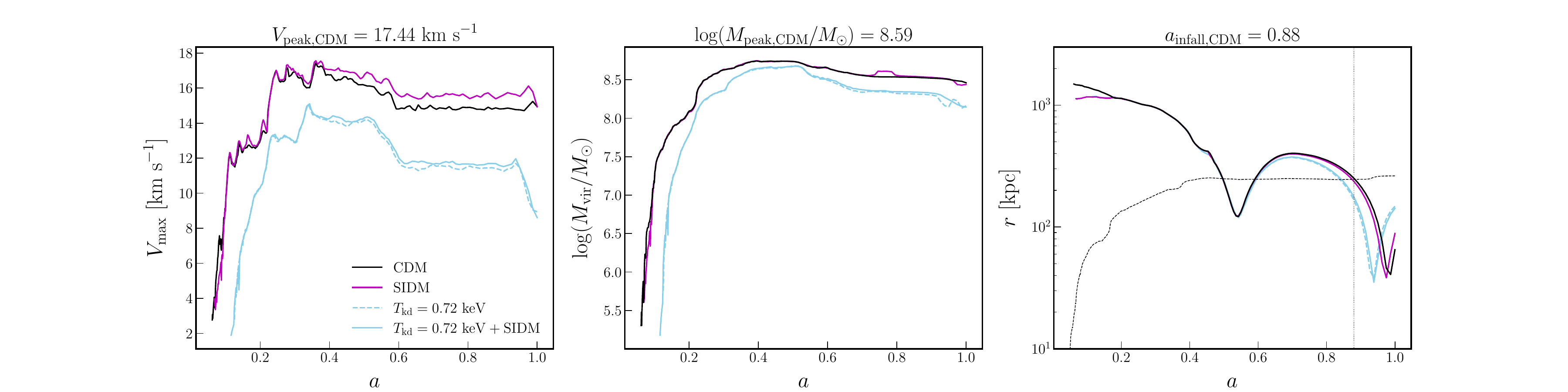}
    \caption{Same as Figure~\ref{fig:matched_halos}, for matched subhalos in our $T_{\mathrm{kd}}=0.72~\mathrm{keV}$ simulations. The high-mass subhalo (top row) is the same as that in Figure~\ref{fig:matched_halos}, while the medium-mass (middle row) and low-mass (bottom row) subhalos are different systems than those shown in the corresponding rows of Figure~\ref{fig:matched_halos}.}
    \label{fig:matched_halos_072}
\end{figure*}

\begin{figure*}[t!]
\hspace{-25mm}
\includegraphics[width=1.25\textwidth]{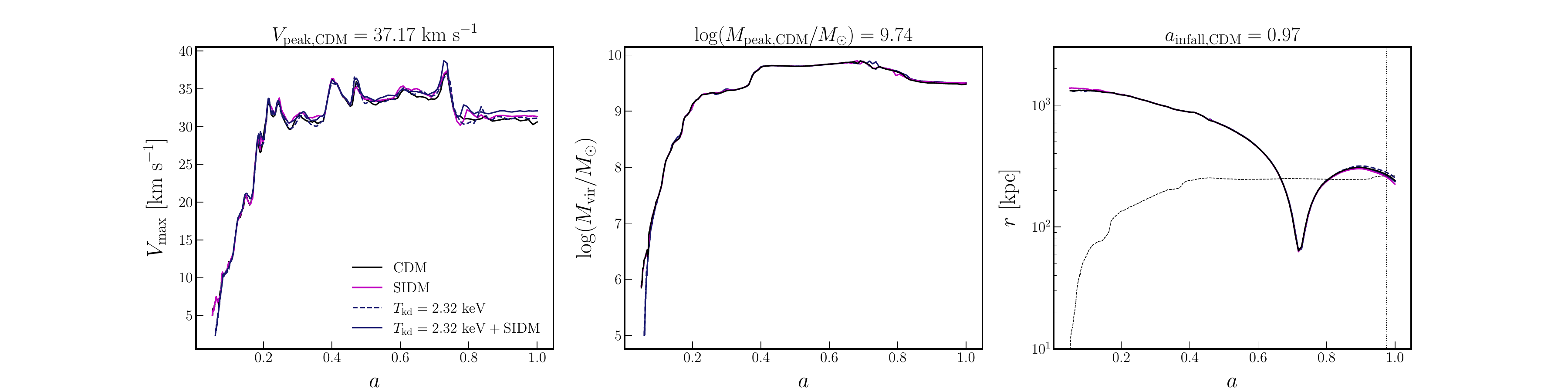} \\
\hspace*{-25mm}
\includegraphics[width=1.25\textwidth]{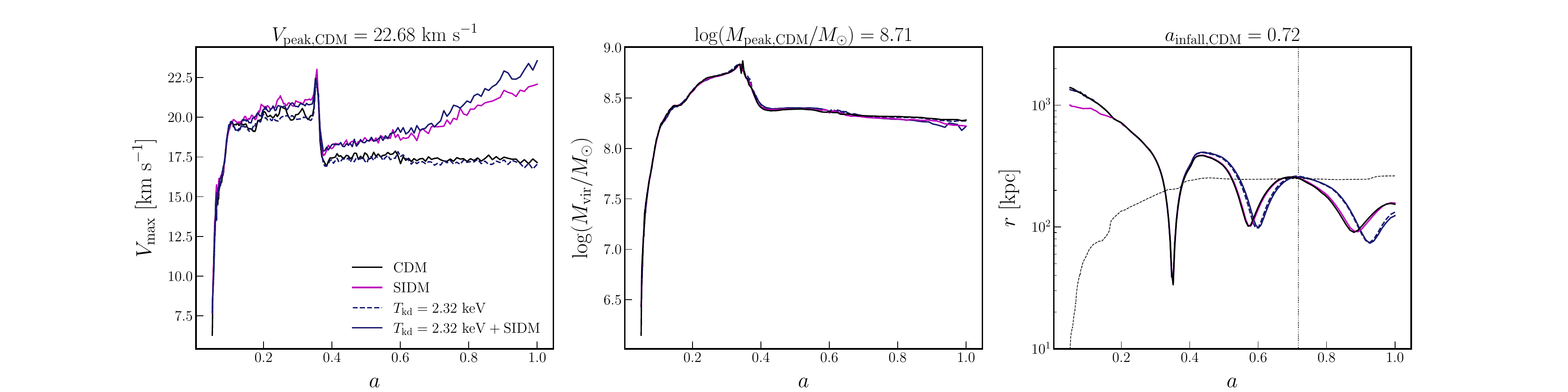} \\
\hspace*{-25mm}
\includegraphics[width=1.25\textwidth]{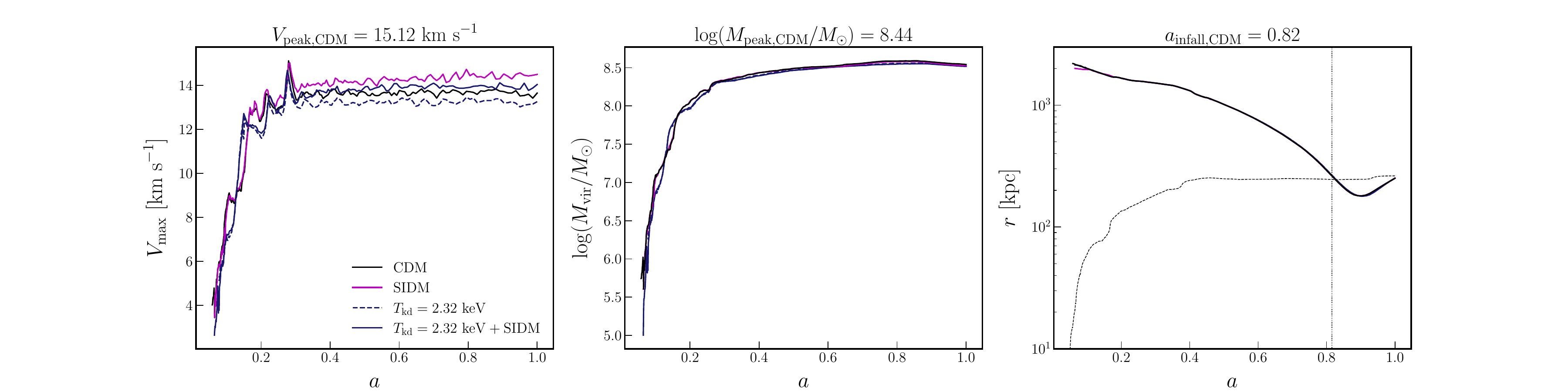}
    \caption{Same as Figure~\ref{fig:matched_halos}, for matched subhalos in our $T_{\mathrm{kd}}=2.32~\mathrm{keV}$ simulations. Each row shows the same system as the corresponding row in Figure~\ref{fig:matched_halos}.}
    \label{fig:matched_halos_232}
\end{figure*}

Figures~\ref{fig:matched_halos_072} and \ref{fig:matched_halos_232} respectively show the evolution of matched subhalos in our $T_{\mathrm{kd}}=0.72$ and $2.32~\mathrm{keV}$ simulations with and without self-interactions, in the same format as Figure~\ref{fig:matched_halos}.

For the $T_{\mathrm{kd}}=0.72~\mathrm{keV}$ case, only the high-mass subhalo (top row) matches the system shown in Figure~\ref{fig:matched_halos}; we cannot identify the medium- and low-mass systems shown in the middle and bottom rows of Figure~\ref{fig:matched_halos} because they do not form in the $T_{\mathrm{kd}}=0.72~\mathrm{keV}$ runs. Instead, the middle and bottom rows of Figure~\ref{fig:matched_halos_072} show subhalos with values of $V_{\mathrm{peak,CDM}}$ comparable to the medium- and low-mass systems in Figure~\ref{fig:matched_halos}. We draw the following conclusions from Figure~\ref{fig:matched_halos_072}.
\begin{itemize}
    \item The high-mass subhalo (top row) has a significantly suppressed growth history relative to its $T_{\mathrm{kd}}=1.46~\mathrm{keV}$ counterpart in Figure~\ref{fig:matched_halos}. This difference is particularly noticeable at late times, and tidal stripping of the $T_{\mathrm{kd}}=0.72~\mathrm{keV}$ subhalo is enhanced near pericenter relative to the CDM and SIDM versions of the subhalo, regardless of whether self-interactions are included.
    \item The moderate-mass subhalo (middle row) is mildly core collapsed in our SIDM simulation, and its accelerated gravothermal evolution appears to follow an early major merger. This subhalo's growth is severely suppressed (and mildly delayed) in the $T_{\mathrm{kd}}=0.72~\mathrm{keV}$ runs; its present-day $V_{\mathrm{max}}$ is slightly enhanced in the WSIDM simulation compared to the $T_{\mathrm{kd}}=0.72~\mathrm{kev}$--only case, but the difference is minor. 
    \item The low-mass subhalo (bottom row) is again mildly core collapsed in our SIDM simulation, while its growth is extremely suppressed and delayed in the $T_{\mathrm{kd}}=0.72~\mathrm{keV}$ runs. In addition, its orbital phase shifts after first pericenter in these runs, relative to the CDM and SIDM versions of the system, which is likely driven by differences in the subhalo's density profile. In Appendix~\ref{sec:host_density}, we show that the host density profile has a similar core in our SIDM and WSIDM simulations and a similar cusp in our CDM and $T_{\mathrm{kd}}$--only simulations; thus, the orbital phase shift cannot be due to differences in the host potential, since the phase is similar in our CDM and SIDM simulations and in our $T_{\mathrm{kd}}=0.72~\mathrm{keV}$--only and WSIDM simulations.
    \end{itemize}

For the $T_{\mathrm{kd}}=2.32~\mathrm{keV}$ case, we recover all three systems shown in Figure~\ref{fig:matched_halos}, and plot them in the corresponding rows of Figure~\ref{fig:matched_halos_232}. We draw the following conclusions from Figure~\ref{fig:matched_halos_232}: 
\begin{itemize}
    \item The high-mass subhalo (top row) behaves very similarly to the $T_{\mathrm{kd}}=1.46~\mathrm{keV}$ case shown in Figure~\ref{fig:matched_halos}; in the $T_{\mathrm{kd}}=2.32~\mathrm{keV}$ WSIDM run, its present-day $V_{\mathrm{max}}$ is slightly enhanced relative to SIDM and to the $T_{\mathrm{kd}}=1.46~\mathrm{keV}$ simulation, but the difference is small.
    \item The moderate-mass subhalo (middle row) has its gravothermal evolution accelerated in the $T_{\mathrm{kd}}=2.32~\mathrm{keV}$ WSIDM simulation, relative to both SIDM and the $T_{\mathrm{kd}}=1.46~\mathrm{keV}$ case. 
    \item The low-mass subhalo (bottom row) reaches a present-day $V_{\mathrm{max}}$ slightly above CDM in the $T_{\mathrm{kd}}=2.32~\mathrm{keV}$ WSIDM simulation, in contrast to its slightly suppressed $V_{\mathrm{max}}$ in the $T_{\mathrm{kd}}=1.46~\mathrm{keV}$ WSIDM case.
    \end{itemize}
    
Intriguingly, all three examples of $T_{\mathrm{kd}}=2.32~\mathrm{keV}$ WSIDM subhalos have slightly enhanced $V_{\mathrm{max}}$ histories relative to both SIDM and our more extreme WSIDM simulations. This contrasts our result in Figure~\ref{fig:tau_0}, where the predicted $\tau_0$ distributions of $T_{\mathrm{kd}}=2.32~\mathrm{keV}$ WSIDM (sub)halos were systematically shifted to lower $\tau_0$ compared to SIDM. However, the $\tau_0$ prediction is based on applying the parametric model to the $T_{\mathrm{kd}}=2.32~\mathrm{keV}$--only simulation, and it is possible that nonlinear effects that are not captured by the parametric model affect (sub)halos' gravothermal evolution in our $T_{\mathrm{kd}}=2.32~\mathrm{keV}$ WSIDM simulation. We leave a detailed exploration of these results to future study.

\section{Dwarf Galaxy Subhalos}
\label{sec:dwarf_alt}

\subsection{Subhalo Density Profiles}
\label{sec:dwarf_dens_alt}

\begin{figure*}[t!]
\centering
\includegraphics[width=0.925\textwidth,trim={0 0.375cm 0 0cm}]{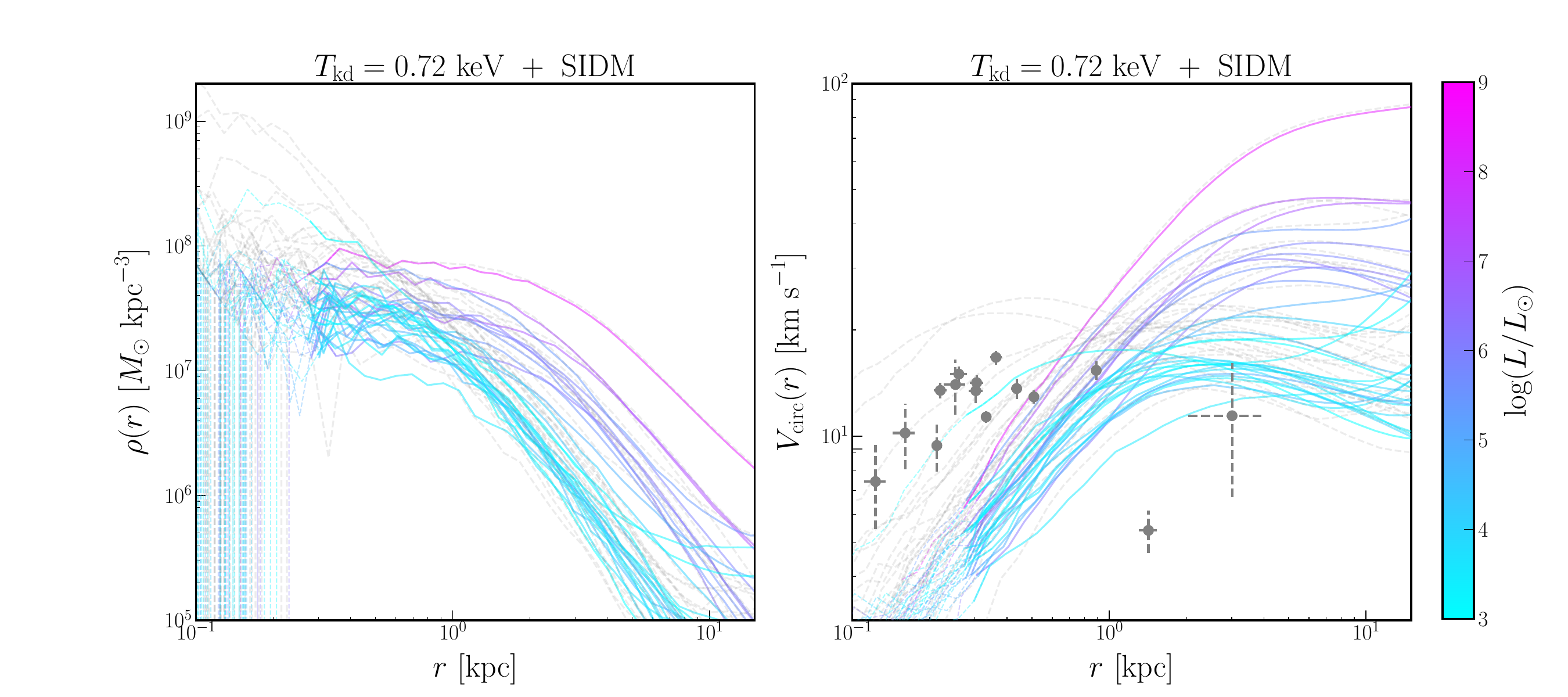} 
\includegraphics[width=0.925\textwidth,trim={0 0.375cm 0 0cm}]{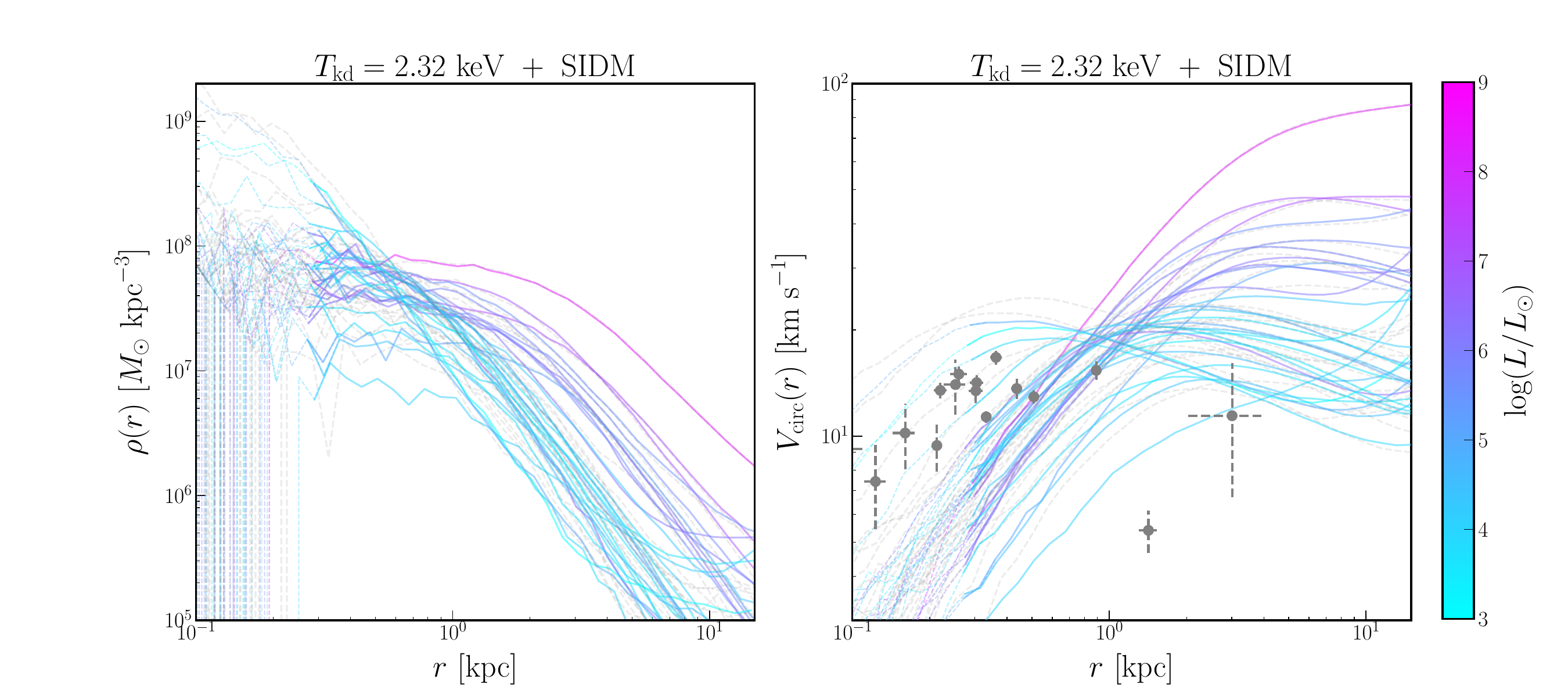} 
    \caption{Same as Figure~\ref{fig:dwarf_density}, for our $T_{\mathrm{kd}}=0.72$ (top) and $2.32~\mathrm{keV}$ (bottom) WSIDM simulations. Faint gray lines in both rows show the SIDM result.}
    \label{fig:dwarf_density_alt}
\end{figure*}

Figure~\ref{fig:dwarf_density_alt} shows subhalo density profiles in our $T_{\mathrm{kd}}=0.72$ and $2.32~\mathrm{keV}$ WSIDM simulations, analogous to Figure~\ref{fig:dwarf_density}; see Section~\ref{sec:dwarf_dens} for a discussion of these results.

\subsection{Central Density--Pericenter Relation}
\label{sec:dwarf_peri_alt}

\begin{figure*}[t!]
\centering
\includegraphics[angle=270,origin=c,width=\textwidth,trim={-0.5cm 0cm 6cm 0cm}]{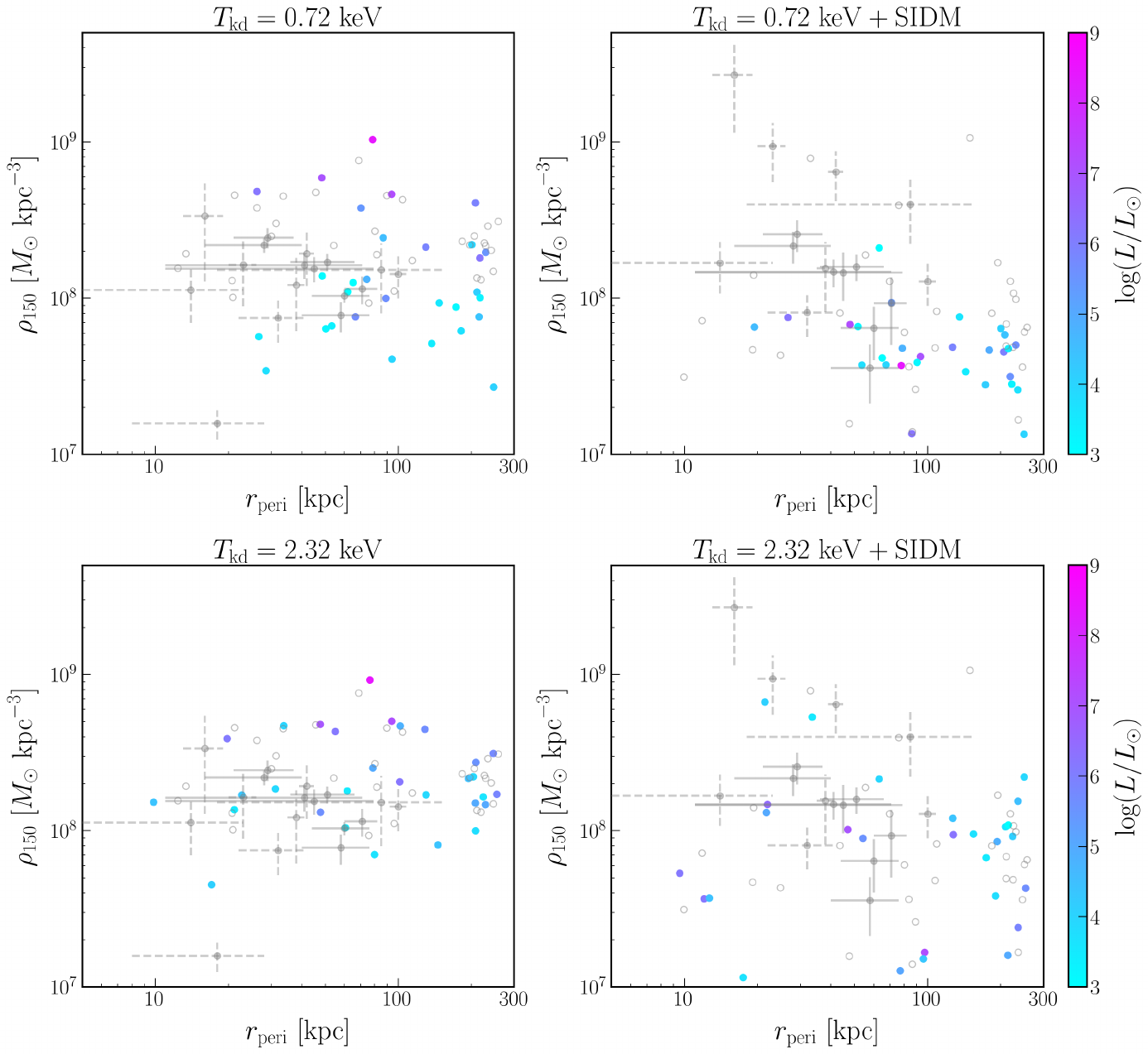}
    \caption{Same as Figure~\ref{fig:rhoperi_rperi}, for our $T_{\mathrm{kd}}=0.72~\mathrm{keV}$--only (top left), $T_{\mathrm{kd}}=0.72~\mathrm{keV}$ WSIDM (top right), $T_{\mathrm{kd}}=2.32~\mathrm{keV}$--only (bottom left), and $T_{\mathrm{kd}}=2.32~\mathrm{keV}$ WSIDM (bottom right) simulations. In the top left panel, open gray circles show the CDM result; in all other panels, they show the SIDM result.}
    \label{fig:rhoperi_rperi_alt}
\end{figure*}

Figure~\ref{fig:rhoperi_rperi_alt} shows subhalo central density--pericenter relations for our $T_{\mathrm{kd}}=0.72$ and $2.32~\mathrm{keV}$ simulations with and without self-interactions; see Section~\ref{sec:dwarf_peri} for a discussion of these results.

\section{Host Halo Density Profiles}
\label{sec:host_density}

\begin{figure}[t!]
\hspace{-5mm}
\includegraphics[trim={0 0cm 0 0cm},width=0.5\textwidth]{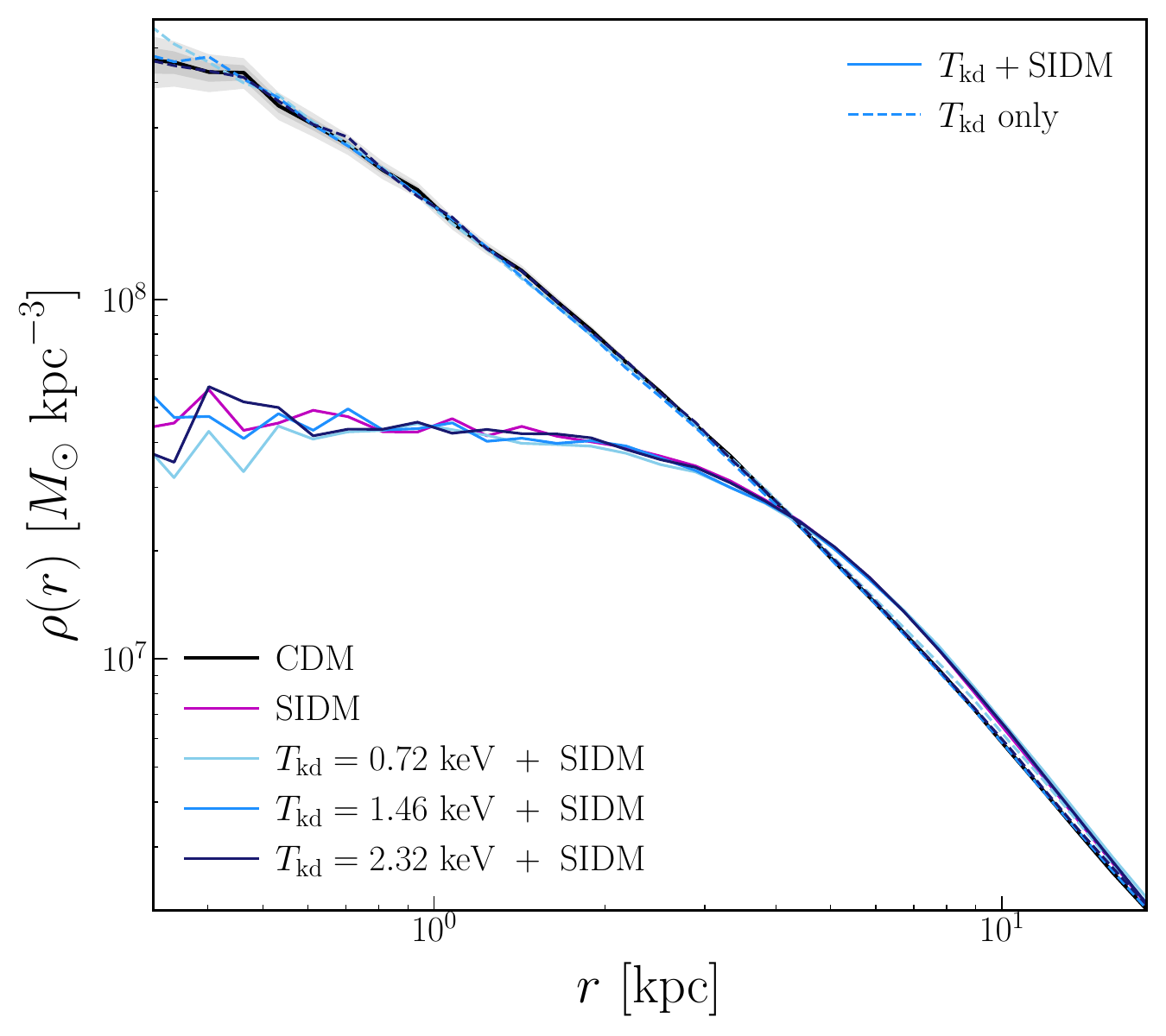}
    \caption{Density profile of our MW--like host halo in CDM (black), SIDM (magenta), and WSIDM models with $T_{\mathrm{kd}}=0.72$, $1.46$, and $2.32~\mathrm{keV}$ (lightest to darkest blue). Corresponding $T_{\mathrm{kd}}$--only simulations are shown by dashed lines. Dark (light) gray bands show $1\sigma$ ($2\sigma$) Poisson uncertainties on the CDM measurement. The minimum radius shown, $2.8\epsilon = 320~\mathrm{pc}$ corresponds to the convergence radius of our simulations. For $r\gtrsim 20~\mathrm{kpc}$, the profiles remain identical out to the virial radius.}
    \label{fig:host_density}
\end{figure}

Figure~\ref{fig:host_density} shows the density profile of the MW--like host halo in our CDM, SIDM, WSIDM, and $T_{\mathrm{kd}}$--only simulations. The host has a prominent kiloparsec-scale density core in WSIDM simulations, independent of the $T_{\mathrm{kd}}$ model and consistent with the core in our SIDM simulation. Note that the SIDM host's core size is similar to that found in \cite{Yang221113768}. Meanwhile, in our $T_{\mathrm{kd}}$--only simulations, the host's density profile is not significantly altered relative to CDM. This is expected because $P(k)$ suppression is negligible on the scale of the host in our $T_{\mathrm{kd}}$ models (see the right panel of Figure~\ref{fig:transfer}), and the host halo mass accretion history is nearly identical in all cases. 

As discussed in Section~\ref{sec:caveats}, we expect these differences between the SIDM and CDM (or WSIDM and $T_{\mathrm{kd}}$--only) density profiles to be erased in the presence of baryons \citep{Kaplinghat13116524,Sameie180109682,Robles190301469,Rose220614830,Correa240309186}.

\end{document}